\documentclass[12pt]{JHEP3}

\usepackage{epsfig}

\def\nth#1{{1 \over #1}}
\def\hf{{1\over 2}}

\def\VEV#1{{\left<{#1}\right>}}
\def\eps{\epsilon}

\def\eq{\begin{equation}}
\def\eeq{\end{equation}}
\def\eqa{\begin{eqnarray}}
\def\eeqa{\end{eqnarray}}
\def\nn{\nonumber}

\def\bd{\begin{displaymath}}
\def\ed{\end{diplaymath}}

\def\Box{ {\,\lower 0.9pt\vbox{\hrule\hbox{\vrule height0.2cm \hskip 0.2cm \vrule height 0.2cm }\hrule}\,}}
\def\lsim{{\ \lower-1.2pt\vbox{\hbox{\rlap{$<$}\lower5pt\vbox{\hbox{$\sim$}}}}\ }}
\def\gsim{{\ \lower-1.2pt\vbox{\hbox{\rlap{$>$}\lower5pt\vbox{\hbox{$\sim$}}}}\ }}

\def\pref#1{(\ref{#1})}

\def\ssubsubsection#1{\vspace{3mm} \noindent \textbf{#1} \\ \vspace{-3mm} \\ \noindent}

\title{SUSY Breaking
and Moduli Stabilization from Fluxes in Gauged 6D Supergravity}
\author{Y. Aghababaie$^1$, C.P. Burgess$^1$, S.L. Parameswaran$^2$ and F. Quevedo$^2$
\\
$^1$ Physics Department, McGill University,  3600 University Street,\\
              Montr\'eal, Qu\'ebec, Canada, H3A 2T8.\\

$^2$ Centre for Mathematical Sciences, DAMTP,
               University of Cambridge,\\
               Cambridge CB3 0WA UK.}

\abstract{We construct the 4D $N=1$ supergravity which describes
the low-energy limit of 6D supergravity compactified on a sphere
with a monopole background {\it \`a la} Salam and Sezgin. This
provides a simple setting sharing the main properties of realistic
string compactifications such as flat 4D spacetime, chiral
fermions and $N=1$ supersymmetry as well as Fayet-Iliopoulos terms
induced by the Green-Schwarz mechanism. The matter content of the
resulting theory is a supersymmetric $SO(3)\times U(1)$ gauge
model with two chiral multiplets, $S$ and $T$. The expectation
value of $T$ is fixed by the classical potential, and $S$
describes a flat direction to all orders in perturbation theory.
We consider possible perturbative corrections to the K\"ahler
potential in inverse powers of $\hbox{Re}\, S$ and $\hbox{Re}\,
T$, and find that under certain circumstances, and when taken
together with low-energy gaugino condensation, these can lift the
degeneracy of the flat direction for Re $S$. The resulting vacuum
breaks supersymmetry at moderately low energies in comparison with
the compactification scale, with positive cosmological constant. It
is argued that the 6D model might itself be obtained from string
compactifications, giving rise to realistic string compactifications
on non Ricci flat manifolds. Possible phenomenological and cosmological
applications are briefly discussed.}

\preprint{McGill-02/34}

\keywords{supersymmetry breaking, string moduli} \preprint

\begin{document}

\section{Introduction}
Realistic string models in 4D share many interesting properties.
They correspond to backgrounds  for which the spacetime is 4D
Minkowski  times an internal 6D manifold, and they have chiral
fermions in 4D  with spectrum close to the Standard Model. They
also have important unsolved issues such as supersymmetry breaking
and the presence of moduli fields such as the dilaton
\cite{review}. Several ideas have been put forward to deal with
these issues, including nonperturbative effects, such as gaugino
condensation \cite{GC}, and the introduction of fluxes of
antisymmetric tensor fields \cite{Flux}. Furthermore
brane/antibrane systems and intersecting branes  have been
considered both to obtain realistic models with broken
supersymmetry \cite{Models} and for the possibility of generating
cosmological inflation \cite{BI,BI2}.

However, there is at the moment not a single model that can
achieve all the successes simultaneously. For instance, to get
inflation it is needed to assume that some of the moduli have been
fixed by an unknown mechanism. Fluxes of Ramond-Ramond fields have
been used to fix some moduli but not all of them. Supersymmetric
models have to face the breaking of supersymmetry, and gaugino
condensation and other nonperturbatively generated superpotentials
usually lead to runaway potentials \cite{GC,recent}.
Non-supersymmetric models such as brane/antibrane systems at
singularities or intersecting brane models tend to be unstable,
with the corresponding scalar potential not under control.

In this article we start an investigation of most of these issues
in a setting that is much simpler than the explicit string
constructions and  yet shares the relevant properties of those
models. The starting point is minimal 6D gauged supergravity
coupled to at least one $U(1)$ vector multiplet. This model was
studied in \cite{SS}, who considered compactification on a
two-sphere stabilized by a nonvanishing magnetic flux through the
sphere for the $U(1)$ gauge field. This compactification was found
to lead to a chiral $N=1$ supersymmetric model in flat 4D
spacetime at scales lower than the compactification scale. We here
reconsider this model and study its consequences in more detail. We
may see this as a first attempt to extract phenomenological
implications to gauged supergravity potentials that are being derived
recently in string theory.
In this article we concentrate only on general issues concerning
the model's low-energy effective action, supersymmetry breaking
and moduli stabilization. We investigate how the introduction of
branes changes the implications of this `bulk' physics in a
subsequent publication.

The model has several stringy properties, such as the presence of
the standard $S$ and $T$ moduli fields of string
compactifications, and we use these to address the issue of
lifting the vacuum degeneracy. Furthermore, the low-energy theory
has a Fayet-Iliopoulos term which is generated by the
Green-Schwarz anomaly-cancelling mechanism. We find this induces a
$D$-term potential in the low-energy theory that can naturally fix
the field $T$, leaving only the $S$ direction flat to all orders
in perturbation theory.

The nonabelian $SO(3)$ symmetry associated to the isometries of
the two-sphere is asymptotically free and so generates a
nonperturbative potential for $S$ through the gaugino condensation
mechanism. Together with the tree-level K\"ahler potential, this
leads to a runaway potential for the dilaton field $S$. This kind
of scenario, with fixed $T$ and with runaway $S$, could provide a
natural way for the model to realize inflation, along the lines
suggested in ref.~\cite{BI2}. We further argue that if certain
perturbative corrections to the K\"ahler potential arise, then the
gaugino-condensation potential need not generate a runaway, and
could be used to stabilize $S$. As is expected on general grounds,
this minimum arises at the margins of what can be computed using
semiclassical methods.

We organize our presentation as follows. The next section gives a
brief review of the relevant features of 6D gauged supergravity,
and its supersymmetric compactification to 4D on a sphere. Section
\pref{S:4DETI} then derives the low-energy 4D supergravity which
describes the low-energy limit of this compactification, as well
as discussing its likely vacuum. The flat directions of the
low-energy theory are the topic of Section
\pref{S:FlatDirections}, where it is shown that all are lifted.
This section shows that some moduli are stabilized at finite
values, while others may be stabilized, or run to infinity,
depending on the details of the corrections to the model's
K\"ahler function. These results, and some of their potential
applications, are discussed in Section \pref{S:Discussion}, where we
also point out that the 6D model itself  may be derived from compactifications of
string theory, either on sphere or $K_3$ compactifications in the
presence of fluxes of antisymmetric tensor fields.

\section{The 6D Salam-Sezgin Model}
\label{S:6DSS}
We begin by recapping the Salam-Sezgin compactification of the
six-dimensional supersymmetric Einstein-Maxwell system
\cite{MS,NS,SS}.

\subsection{The Model}
The field content of the theory consists of a supergravity
multiplet -- which comprises a metric ($g_{MN}$), antisymmetric
Kalb-Ramond field ($B_{MN}$), dilaton ($\phi$), gravitino
($\psi_M$) and dilatino ($\chi$) -- coupled to a $U(1)$ gauge
multiplet -- containing a gauge potential ($A_M$) and gaugino
($\lambda$).

The fermions are all complex Weyl spinors -- satisfying $\Gamma_7
\psi_M = \psi_M$, $\Gamma_7 \lambda = \lambda$ and $\Gamma_7 \chi
= - \chi$ -- and they all transform under the $U(1)$ gauge
symmetry. For instance, the gravitino covariant derivative is
\begin{equation} \label{E:covderiv}
    D_{M}\psi_{N} = \left(\partial_{M} -
    \frac{1}{4}{\omega_{M}}^{AB}\Gamma_{AB} - igA_{M}\right)\psi_{N} ,
\end{equation}
where ${\omega_{M}}^{AB}$ denotes the spin connection. Here $g$
denotes the 6D $U(1)$ gauge coupling, which in fundamental units
($\hbar = c= 1$) has the dimension (mass)${}^{-1}$.

The field strength for $B_{MN}$ contains the usual supergravity
Chern-Simons contribution
\begin{equation}
    \label{E:CSterm}
    G_{MNP} = \partial_{M}B_{NP} + F_{MN}A_{P} + \hbox{(cyclic
    permutations)} \, ,
\end{equation}
where $F_{MN} = \partial_M A_N - \partial_N A_M$ is the usual
abelian gauge field strength. Notice that the appearance of $A_M$
in this equation implies $B_{MN}$ must also transform under the
$U(1)$ gauge transformations, since invariance of $G_{MNP}$
requires
\eq \label{E:Btransf} \delta A_M =
\partial_M \omega , \qquad \delta
B_{MN} = - \omega \, F_{MN} \, . \eeq
This transformation allows the gauge anomalies due to the chiral
fermions to be cancelled by a Green-Schwarz mechanism, as must
happen if this supergravity emerges as a low-energy
compactification of string theory (more about this below).

The bosonic part of the classical 6D supergravity action
is:\footnote{Our metric is `mostly plus' and like all
right-thinking people we follow Weinberg's curvature conventions
\cite{GandC}.}
\begin{equation}
\label{E:Baction}
    e^{-1} {\cal L}_B = -\, \frac{1}{2 } R - \frac{1}{2 }
    \partial_{M} \phi \, \partial^M\phi  -
    \frac{e^{-2\phi}}{12} \; G_{MNP}G^{MNP} -
    \frac{e^{-\phi}}{4} \; F_{MN}F^{MN} - {2 g^2 e^\phi},
\end{equation}
where we choose units for which the 6D Planck mass is unity:
$\kappa_6^2 = 8 \pi G_6 = 1$. As usual $e = |\det {e_M}^A| =
\sqrt{-\det g_{MN}}$.

The part of the action which is bilinear in the fermions is
\eqa \label{E:Faction}
    e^{-1} {\cal L}_F &=& -\bar{\psi}_{M}\Gamma^{MNP}D_{N}\psi_{P} -
    \bar{\chi}\Gamma^{M}D_{M}\chi - \bar{\lambda}\Gamma^{M}D_{M}\lambda \nn\\
    &&-\frac{1}{2} \partial_{M}\phi
    \left(\bar{\chi}\Gamma^{N}\Gamma^{M}\psi_{N} +
    \bar{\psi}_{N}\Gamma^{M}\Gamma^{N}\chi\right) \nn\\
    &&+\frac{e^{-\phi}}{12 \sqrt{2}} \; G_{MNP}
    (-\bar{\psi}^{R}\Gamma_{[R}\Gamma^{MNP}\Gamma_{S]}\psi^{S}
    +\bar{\psi}_{R}\Gamma^{MNP}\Gamma^{R}\chi\\
    &&\phantom{00000000000000000}-\bar{\chi}\Gamma^{R}\Gamma^{MNP}\psi_{R}
    + \bar{\chi}\Gamma^{MNP}\chi - \
    \bar{\lambda}\Gamma^{MNP}\lambda)\nn\\
    &&- \; \frac{e^{-\phi/2}}{4} \; F_{MN}
    \left(\bar{\psi}_{Q}\Gamma^{MN}\Gamma^{Q}\lambda
    + \bar{\lambda}\Gamma^{Q}\Gamma^{MN}\psi_{Q} -
    \bar{\chi}\Gamma^{MN}\lambda +
    \bar{\lambda}\Gamma^{MN}\chi\right)\nn\\
    &&+ i g e^{\phi/2} \left(\bar{\psi}_{M}\Gamma^{M}\lambda+
    \bar{\lambda}\Gamma^{M}\psi_{M}+ \bar{\chi}\lambda - \bar{\lambda}\chi
    \right),\nn
\eeqa
where the completely antisymmetric products of Dirac matrices are
defined by $\Gamma_{MN} = \frac12 \left(\Gamma_M \Gamma_N -
\Gamma_N \Gamma_M \right)$, $\Gamma_{MNP} = \frac16 \,
\left(\Gamma_M \Gamma_N \Gamma_P \pm \hbox{permutations} \right)$
and so on.

The 6D supersymmetry transformations which preserve the form of
this action are
\eqa \label{E:susy}
    \delta e^{A}_{M} &=& {1 \over \sqrt{2}} \; \Bigl(
    \bar{\epsilon}\Gamma^{A}\psi_{M} - \bar{\psi}_{M}\Gamma^{A}\epsilon
    \Bigr) \nn\\
    \delta\phi &=& - \; {1\over \sqrt2 }
    \Bigl( \bar{\epsilon}\chi + \bar{\chi}\epsilon \Bigr) \nn \\
    \delta B_{MN} &=& \sqrt{2}  \, A_{[M}\delta A_{N]}
    + \frac{e^\phi}{2} \,
    \Bigl( \bar{\epsilon}\Gamma_{M}\psi_{N}-\bar{\psi}_{N}\Gamma_{M}\epsilon \nn \\
    && \qquad - \bar{\epsilon}\Gamma_{N}\psi_{M}
    +\bar{\psi}_{M}\Gamma_{N}\epsilon -
    \bar{\epsilon}\Gamma_{MN}\chi +
    \bar{\chi}\Gamma_{MN}\epsilon\Bigr)\\
    \delta\chi &=&
    \frac{1}{\sqrt2}\; \partial_{M}\phi \; \Gamma^{M}\epsilon
    + \frac{e^{-\phi}}{12} G_{MNP} \; \Gamma^{MNP}\epsilon \nn\\
    \delta\psi_{M} &=& \sqrt{2} \; D_{M}\epsilon +
    \frac{e^{-\phi}}{24} \; G_{PQR} \; \Gamma^{PQR}\Gamma_{M}\epsilon \nn\\
    \delta A_{M} &=&
    \frac{1}{\sqrt{2}} \Bigl(\bar{\epsilon}\Gamma_{M}\lambda
    - \bar{\lambda}\Gamma_{M}\epsilon\Bigr) e^{\phi/2} \nn\\
    \delta\lambda &=&
    \frac{e^{-\phi/2}}{4} \; F_{MN} \; \Gamma^{MN}\epsilon
    - \frac{i}{\sqrt{2}} \; g \, e^{\phi/2} \, \epsilon \,
    , \nn
\eeqa
where the supersymmetry parameter is complex and Weyl: $\Gamma_7
\epsilon = \epsilon$.

\ssubsubsection{Global and Approximate Symmetries}
Besides supersymmetry, the $U(1)$ gauge symmetry and the
Kalb-Ramond symmetry, $\delta B = d\Lambda$, the model also has a
few other symmetries (and approximate symmetries) which are useful
to enumerate here for later convenience.

First, if the background spacetime admits an harmonic 2-form,
$\Omega$, then because $B$ only enters the action through $dB$ the
background has a global symmetry $\delta B = c \, \Omega$, where
$c$ is the constant symmetry parameter. Because $\Omega \ne
d\Lambda$ for any globally-defined 1-form this symmetry can be
regarded as independent of the Kalb-Ramond gauge symmetry.

Second, the field equations obtained from this action have a
classical symmetry under the following constant rescaling of the
fields:
\eq \label{E:ttransf} g_{MN} \to \sigma \, g_{MN}, \qquad e^\phi
\to e^\phi/\sigma, \qquad \psi_M \to \sigma^{1/4} \, \psi_M,
\qquad \chi \to \chi / \sigma^{1/4}, \qquad \lambda \to \lambda /
\sigma^{1/4} , \eeq
with no other fields transforming. This is just a symmetry of the
equations of motion, rather than a {\it bona fide} symmetry
because it does not leave the action invariant, but rather
rescales it according to ${\cal L} \to \sigma^2 {\cal L}$. This
symmetry reflects the possibility of performing redefinitions to
write the Lagrangian density as ${\cal L} = e^{-2\phi} \, {\cal
L}_{\rm inv}$, with ${\cal L}_{\rm inv}$ a function only of the
invariant `string-frame' quantities $g^s_{MN} = e^\phi \, g_{MN}$,
$\psi^s_M = e^{\phi/4}\, \psi_M$, $\chi^s = e^{-\phi/4}\, \chi$,
$\lambda^s = e^{-\phi/4}\, \lambda$ and $\partial_M \phi$. In
general, since only the dilaton transforms under the scaling
symmetry in the string frame, the $\ell$-string-loop contribution
to the action scales as ${\cal L}_\ell \to \sigma^{2 - 2\ell} \,
{\cal L}_\ell$.

\subsection{Anomaly Cancellation}
As mentioned above, the fermion content of the 6D model as
described so far has anomalies, which must be cancelled if the
theory is to make physical sense. In particular, they must cancel
if the model is to be considered as the low-energy limit of an
underlying consistent theory, such as string theory, at still
higher energies. The anomaly cancellation conditions for 6D
supergravity coupled to a single tensor- plus $n_V$ vector- and
$n_H$ hyper-multiplets are well understood, so we simply summarize
here several features which are used below.

These anomaly-cancelling conditions are significant for two
separate reasons. First, they provide new nontrivial constraints
on the kinds of particles which must appear in the 6D theory. In
particular they require the existence of many more 6D matter
supermultiplets than are considered above. Second, they show the
necessity for specific higher-derivative corrections to the above
supergravity lagrangian without reference to any specific, more
microscopic, theory like string theory. As we shall see, the
neglect of these corrections when using the above 6D supergravity
lagrangian ultimately requires for consistency the conditions
\eq 1/r^2 \ll e^\phi \ll 1 \, , \eeq
with $r$ defined in terms of the volume of the extra dimensions
using the Einstein-frame metric.

\ssubsubsection{The 6D Green-Schwarz Mechanism}
Six-dimensional anomalies are described by an 8-form constructed
from the gauge and gravitational field strengths \cite{AGW}. In
order for anomalies to be cancelled by a Green-Schwarz type
mechanism \cite{GSAC} -- involving the shifting of a bosonic field
in the theory -- this anomaly form must factorize into the wedge
product of pairs of lower-dimension forms.\footnote{If several
bosonic fields are involved then the anomaly 8-form can be the sum
of such products, one for each of the bosonic fields.} In general
this imposes a strong set of conditions in six dimensions, some
features of which we now summarize \cite{RDSSS, 6DAC}.

To describe the anomaly cancellation conditions we must first
generalize the above field content to potentially include $n_V$
gauge multiplets, as well as $n_H$ matter `hyper-' multiplets which
involve 6D scalars and fermions whose helicity satisfies $\Gamma_7
= -1$. 6D supersymmetry requires the scalars within these
hypermultiplets to take values in a quaternionic manifold, and
precludes them from appearing in the gauge kinetic terms or in the
kinetic term for the dilaton field $\phi$ \cite{janber}.

A necessary condition for the factorizing of the anomaly 8-form is
the vanishing of the coefficient of the tr$(R^4)$ term. With $n_V$
gauge multiplets and $n_H$ hypermultiplets, this is assured by the
condition $n_H = n_V + 244$, which determines the number of
hyper-multiplets in terms of the dimension of the 6D gauge group,
$n_V=\hbox{dim} G$ \cite{RDSSS}. For simple gauge groups whose
quartic casimir invariant is linearly independent of those at
lower orders, there is another condition which amounts to
requiring the vanishing of the tr$(F^4)$ term.

If the only gauge group is the $U(1)$ considered above, we have
$n_V =1$ and so $n_H = 245$ hyper-multiplets are required to
cancel anomalies. (For our purposes more general gauge groups may
also be possible, provided that the compactification solution we
introduce in the next section remains a solution to the equations
of motion.) In this case anomaly cancellation through a shift in
the field $B_{MN}$ requires the anomaly 8-form to be
\eq \label{E:Anomaly}
  I_8 =  k \Bigl(\hbox{tr}\, R^2 - v  F^2 \Bigr) \;
  \Bigl( \hbox{tr}\,R^2 - \tilde v F^2 \Bigr) \, ,
\eeq
where the trace is over the fundamental representation of
$SO(5,1)$ and multiplication represents the wedge product. Here
the numbers $k$, $v$ and $\tilde v$ are calculable given the
precise fermion content of the 6D theory.

An anomaly of this form may be cancelled by adding the
Green-Schwarz term
\eq \label{E:AnomCan}
    {\cal L}_{\rm anom} = -k v B \Bigl( \hbox{tr}\, R^2 - \tilde v
    F^2\Bigr),
\eeq
provided the transformation rule, eq.~\pref{E:Btransf}, for
$B_{MN}$ is modified to $\delta B = - \omega \, F + \alpha_L/v$
where $d\alpha_L = \delta \omega_L$ gives the transformation
property of the gravitational Chern-Simons form, $\omega_L$, which
is in turn defined by the condition $d\omega_L = \hbox{tr}\, R^2$.
Invariance of the field strength, $G_{MNP}$, then requires its
definition be modified to $G = dB + AF -  \omega_L/v$.

The anomaly-cancelling term and the modifications to $G$ are
linked by supersymmetry to one another, and to other
higher-derivative terms in the 6D action beyond those described
above. For instance in the Einstein frame the $U(1)$ gauge kinetic
functions get modified to \cite{Sagnotti,dmw}
\eq \label{E:GkinCorr}
    - \, \frac14 \, \left( e^{-\phi} + {\tilde v \over v} \, e^\phi \right)
    F_{MN} F^{MN} \, .
\eeq
As we see in more detail once the compactification to four
dimensions is described below, all of these new terms are
suppressed relative to the ones discussed in the previous sections
in the sense that they involve higher powers of either $1/(e^\phi
r^2)$ or $e^\phi/r^2$ (or both).

\subsection{The Compactification}
The equations of motion for the bosonic fields which follow from
the action, eq.~\pref{E:Baction}, are:
\eqa \label{E:Beom} &&\Box \, \phi + \frac16 \, e^{-2 \phi} \,
G_{MNP} \, G^{MNP} + \frac14 \, e^{-\phi} \; F_{MN} F^{MN} - 2 g^2
e^\phi = 0 \nn\\
&&D_M \Bigl( e^{-2\phi} \, G^{MNP} \Bigr) = 0  \\
&&D_M \Bigl( e^{-\phi} \, F^{MN} \Bigr) + e^{-2\phi} \, G^{MNP} \,
F_{MP} = 0 \nn \\
&&R_{MN} + \partial_M\phi \, \partial_N \phi + \frac12 \,
e^{-2\phi} \, G_{MPQ} \, {G_N}^{PQ} + e^{-\phi} \, F_{MP} {F_N}^P
+ \frac12 \,  (\Box \phi )\, g_{MN} = 0 . \nn \eeqa

The compactification is found by searching for a solution to these
equations which distinguishes four of the dimensions -- $x^\mu,
\mu = 0,1,2,3$ -- from the other two -- $y^m, m=4,5$. The
Salam-Sezgin solution is obtained by constructing this solution
subject to the symmetry ansatz that the spacetime be separately
maximally symmetric in the first four and last two dimensions.
This leads to the following Freund-Rubin-type ansatz \cite{FR} for
the solution: $\phi =$ constant and
\eq \label{E:FRansatz}
    {g}_{MN} = \pmatrix{
    {g}_{\mu\nu}(x) & 0 \cr 0 & {g}_{mn}(y) \cr}
    \qquad \hbox{and} \qquad
    {F}_{MN} = \pmatrix{0 & 0 \cr 0 &
    {F}_{mn}(y) \cr}  ,
\eeq
where ${g}_{\mu\nu}$ is a maximally-symmetric Lorentzian metric
({\it i.e.} de Sitter, anti-de Sitter or flat space), and
${g}_{mn}$ is the metric on the two-sphere, $S_2$. Maximal
symmetry implies the gauge field strength is proportional to the
sphere's volume form, ${\epsilon}_{mn}$, and so
\eq {F}_{mn} = f \; {{\epsilon}_{mn}} \, , \eeq
where $f$ is a constant. All other fields vanish.

The gauge potential, ${A}_m$, which gives rise to this field
strength is the potential of a magnetic monopole. As such, it is
subject to the condition that the total magnetic flux through the
sphere is quantized: $g \int_{S_2} B \, d^2y = 2 \pi n$, with $n =
0,\pm 1,..$. This requires the normalization constant, $f$, to be:
\eq \label{E:fquant}
    f = {n \over 2 \, g \, {r}^2} \eeq
where ${r}$ is the radius of the sphere.

As is easily verified, the above ansatz solves the field equations
provided that the following three conditions are satisfied:
${R}_{\mu\nu} = 0$, ${F}_{mn} {F}^{mn} = 8\, g^2 e^{2{\phi}}$ and
${R}_{mn} = - \, e^{-{\phi}} \, {F}_{mp} \, {{F}_n}^p = - f^2
e^{-{\phi}} \, {g}_{mn}$.\footnote{In reference \cite{RDSSS} the
authors construct a similar solution by embedding the monopole in
an $E_6$, also achieving flat four-dimensional space.  The
solution in~\cite{RDSSS} breaks supersymmetry.} These imply the
four dimensional spacetime is flat, the monopole number is $n =
\pm 1$ and the sphere's radius is related to ${\phi}$ by
\eq \label{E:phircond}  e^{\phi} \, {r}^2 = {1 \over 4 g^2} \, .
\eeq

Useful intuition about this result can be obtained by constructing
the scalar potential for $r$ and $\phi$ which is obtained by
substituting our assumed background solution into the classical
action. One finds in this way three contributions, coming from the
Einstein-Hilbert term, the Maxwell kinetic term for $A_m$ and the
explicit dilaton potential. In order to eliminate mixing between
these scalars and the fluctuations of the 4D metric, it is
necessary to perform a Weyl rescaling to ensure the 4D
Einstein-Hilbert action remains $r$-independent. We take,
then:\footnote{We implicitly change units when performing this
rescaling, switching to the choice $\kappa_4^2 = 8 \pi G_4 = 1$,
rather than the same condition for the 6D quantity, $\kappa_6^2$.}

\eq \label{E:ganz}
    g_{MN} = \pmatrix{r^{-2} \, g_{\mu\nu} & 0 \cr
    0 & r^2 g_{mn} \cr} \eeq
and find the following potential:
\eq \label{E:rpot}
    V = - \; \left.  {{\cal L}_B \over e_4}
    \right|_{\rm no \; derivatives} = {2 \, g^2 \, e^\phi \over r^2} \;
    \left( 1 - \; {1 \over 4 g^2 \, e^\phi r^2} \right)^2 , \eeq
where $e_4 = \sqrt{-\det g_{\mu\nu}}$. {}From this we see how
eq.~\pref{E:phircond} emerges as the minimum of the scalar
potential for $r$ and $\phi$. Because this potential is minimized
at $V=0$ we also see why the 4D metric must be flat. Finally, we
see that the combination $e^\phi/r^2$ parameterizes a flat
direction, since its potential vanishes identically once $e^\phi
r^2 = 1/4 g^2$ has been chosen.

The existence of the flat direction parameterized by $r^2/e^\phi$
may be also inferred from the scaling symmetry of the supergravity
equations of motion, eq.~\pref{E:ttransf}. Since $s:=r^2/e^\phi$
transforms under this transformation while $t:=e^\phi \, r^2$ does
not, $s$ plays the role of the dilaton for this symmetry. Since
this scaling transformation is only a symmetry of the classical
equations, and not of the action, the potential for $s$ need not
be exactly flat if we go beyond the classical approximation when
computing the low-energy theory.

In the present case it happens that the flat direction with
vanishing 4D cosmological constant is not lifted order-by-order in
perturbation theory, as may be seen because the solution leaves
one 4D supersymmetry unbroken. This may be seen by substituting
the solution into the right-hand-side of eqs.~\pref{E:susy} and
checking that the result vanishes for a supersymmetry parameter
which is independent of the 2D coordinates, $y^m$. Equivalently,
spinors on $S_2$ which are constants are Killing spinors for this
solution. Their existence is a consequence of the choice $n = \pm
1$ for the monopole number, since this ensures the cancellation of
the gauge and spin connections in the covariant derivative, $D_\mu
\varepsilon$ \cite{SS}.

Some consistency conditions need be borne in mind if we regard
this field configuration as a low-energy solution in string
theory. In this case the approximation of weak string coupling
requires we take $e^\phi \ll 1$ and the approximation of using a
low-energy field theory similarly requires $r \gg 1$. Both of
these requirements imply small values for the combination
$e^\phi/r^2$.

\subsection{Low-Energy Fluctuations}
Fluctuations about this background may be organized into
four-dimensional fields according to the usual Kaluza-Klein
procedure, with the generic mode having a mass which is at least
of order $1/r$. We wish to identify the effective four-dimensional
theory which governs the physics below this scale.

\ssubsubsection{Symmetries}
As a preliminary to the identification of the light particle
content of the 4D theory, we first identify how the background
fields transform under the model's symmetries. Given the
background fields of present interest --- $g_{\mu\nu}, g_{mn}$ and
$F_{mn}$ --- these are:
\begin{itemize}
\item Unbroken 4D Poincar\'e invariance, as given by the
isometries of 4D Minkowski space:
\eq \VEV{\delta g_{\mu\nu}} = \nabla_\mu \xi_\nu + \nabla_\nu
\xi_\mu = 0. \eeq
These symmetries ensure the masslessness of the 4D graviton.
\item Unbroken $SO(3)$ invariance from the isometries of the internal
metric,
\eq
\VEV{\delta g_{mn}} = \nabla_m Z_n + \nabla_n Z_m = 0 \eeq
which ensures the masslessness of three 4D spin-one particles.
\item Broken local $U(1)$ invariance, broken because of the
transformation
\eq \VEV{\delta B} = \omega \VEV{F} . \eeq
From this we draw two conclusions. First, the 4D gauge field
$A_\mu$ is not exactly massless. Second, we see that in the
Kaluza-Klein expansion $B_{mn}(x,y) = b \, \eps_{mn} + \dots$
(where $\eps_{mn}$ is the 2D volume form) the field $b$ mixes with
the Goldstone Boson for the $U(1)$ gauge symmetry breaking.
\item Unbroken 4D Kalb-Ramond symmetry
\eq \VEV{\delta B_{\mu \nu} } = \partial_{\mu} v_{\nu} -
\partial_\nu v_\mu = 0 \eeq
for constant $v_\mu$.
\item Because the 2-sphere's volume form, $\eps_{mn}$, is harmonic, the
action has a global symmetry, $\delta B_{mn} = c \, \eps_{mn}$,
and this is superficially broken by the background, since
$\VEV{\delta B_{mn}} \ne 0$. However because $\VEV{F_{mn}} = f \,
\eps_{mn}$, there is a linear combination of this global symmetry
and the $U(1)$ gauge symmetry which is unbroken by the background
fields:
\eq \VEV{B_{mn}} = c \, \eps_{mn} + \omega \, \VEV{F_{mn}} = 0,
\eeq
provided $c = - f \, \omega$. This shows that in the Kaluza-Klein
expansion $B_{mn} = b \, \eps_{mn}$, the field $b$ becomes
massless in the limit when either $f$ or the $U(1)$ gauge coupling
vanish.
\end{itemize}

\ssubsubsection{Particle Content}
On symmetry grounds we expect the following bosonic particle
content of the effective theory well below the scale $1/r$. (The
corresponding fields are also given up to mixing due to the
nonzero background flux $\VEV{F_{mn}}$).
\vspace{3mm}
\begin{center}
\begin{tabular}{ccc}
No. & Spin & Field \\
1 & 2 & $g_{\mu\nu}(x)$ \\
4 & 1 & $A^a_{\mu}(x)$, 3 combinations of $g_{m\mu}(x,y)$  \\
4 & 0 & $B_{\mu\nu}(x)$, $\phi(x)$, $r(x)$, $B_{mn} = b(x) \,
\epsilon_{mn}/e_2$
\end{tabular}
\end{center}
\vspace{3mm}
where we count here real scalar fields.

This counting arises as follows:
\begin{itemize}
\item
The massless spin-2 particle follows as the gauge particle for the
unbroken 4D Lorentz invariance of the background metric.
\item
The three massless spin-1 particles which arise as combinations
of $g_{\mu n}$ are the gauge bosons for the $SO(3)$ group of
isometries of the 2-sphere.
\item
The field $B_{\mu\nu}$ dualizes to a massless scalar, $a$,
according to the definition $\partial_\mu a = e^{-2 \phi}
\epsilon_{\mu\nu\lambda\rho} G^{\nu\lambda\rho}/e_4$. The symmetry
$a \to a + \hbox{constant}$ can be broken by anomalies, which can
arise after dualization due to the appearance of the Chern Simons
terms in the field strength $G_{\mu\nu\lambda}$.

\item
The fields $b$ and $A_\mu$ are {\it not} massless. The gauge field
is not massless because the background field $F_{mn} \ne 0$ breaks
the $U(1)$ gauge symmetry, with an expectation value which is of
order $f = \pm 1/(2 g r^2)$ in size. The covariant derivative for
$b$ is $\partial_\mu b + f A_\mu$, indicating that $b$ is the
Goldstone boson which is eaten by the gauge boson.\footnote{The
possibility that these fields get a mass appears to have been
missed in ref.~\cite{SS}, but was recognised in ref.~\cite{RDSSS}
in a similar context.}
\item
The combination $t := e^{\phi} r^2$ is also not massless since it
is not a modulus of the background configuration, being fixed by
the condition \pref{E:phircond}.\footnote{The possibility that
fluxes could freeze geometric moduli has been noted previously
in~\cite{sethi}.} We shall see that this scalar's mass is also
suppressed by powers of $e^\phi$ and so can appear in the
low-energy theory below $1/r$.
\item
The orthogonal combination $s := r^2 e^{-\phi}$ is massless in the
classical approximation, as we saw from the scalar potential,
eq.~\pref{E:rpot}.
\end{itemize}

\ssubsubsection{Light Boson Masses}
To see why the masses of the fields $A_\mu$ and $t = r^2 e^\phi$
are suppressed by powers of $e^\phi$, we must compute their
kinetic terms in addition to their mass terms. For instance, for
the gauge field, $A_\mu$, the mass term arises from the square of
the term $F_{mn} A_\mu$ which appears in the Kalb-Ramond kinetic
term. Keeping in mind the Weyl rescaling of the 4D metric this
gives a mass term of order
\eq \label{E:Amass}
   { {\cal L}_{\rm mass} \over e_4 \, e_2} = -\; \frac{1}{4} e^{- 2\phi}
   \; G_{mn\mu} \, G^{mn\mu} \sim  e^{-2 \phi} F_{mn} F^{mn}
    A_\mu A^\mu \sim {e^{-2\phi} \over g^2 r^2} \; A_\mu A^\mu
   \, , \eeq
where $e_2 = \sqrt{\det g_{mn}}$. By contrast, the kinetic term is
\eq \label{E:Akin}
    {{\cal L}_{\rm kin} \over e_4 \, e_2} = - \; \frac14 \, e^{-\phi}
    \, F_{\mu\nu} F^{\mu\nu} \sim r^4 e^{-\phi} \, F_{\mu\nu}
    F^{\mu\nu}. \eeq
Comparing these gives a gauge boson mass of order:
\eq \label{E:Am}
    m^2_A \sim {e^{-\phi} \over g^2 r^6} \sim
{1\over g^2}\,{1\over s\,t^2}
 \sim {g^2 \over s}, \eeq
where we have used the condition $g^2 \, r^2 \, e^\phi = O(1)$, $t
= r^2\, e^\phi$ and $s = r^2/e^\phi$.

The mass for $t$ is found in an identical way. The kinetic term
for $r$ arises from substituting the ansatz, eq.~\pref{E:ganz},
into the 6D Einstein-Hilbert term. Together with the explicit
$\phi$ kinetic term this leads to the following kinetic terms for
$t$ and $s$:
\eqa \label{E:rphikin}
    {\cal L}_{\rm kin} &=& - g^{\mu\nu} \left[2\; {\partial_\mu r \,
    \partial_\nu r \over r^2} + \frac12 \; \partial_\mu \phi
    \, \partial_\nu \phi  \right] \nonumber \\
    &=& - \frac14 \; g^{\mu\nu} \left[ {\partial_\mu s \, \partial_\mu
    s \over s^2 } + {\partial_\mu t \, \partial_\nu t \over t^2 }
    \right]. \eeqa
In terms of $s$ and $t$ the potential, eq.~\pref{E:rpot}, becomes
\eq \label{E:stpot}
     V = {2 \, g^2  \over s} \;
    \left( 1 - \; {1 \over 4 g^2 \, t} \right)^2, \eeq
and so $d^2 V/dt^2 \Bigr|_{\rm min} = 4\, g^2/(s t^2)$. Comparing
with the kinetic term gives a mass which is of the same order as
was found above for $m_A^2$:
\eq m_t^2 \sim {g^2 \over s} \sim {g^2 e^\phi \over r^2} \, . \eeq

For the purposes of comparison, it is worth also recording here
the generic size of Kaluza-Klein masses. For instance, given a
massless 6D scalar field, $\Phi(x,y)$, and keeping in mind the
metric rescaling, eq.~\pref{E:ganz}, we may write
\eq
    g^{MN} \nabla_M \nabla_N \Phi = \left( r^2 \, g^{\mu\nu} \nabla_\mu
    \nabla_\nu + \frac{1}{r^2} \, g^{mn} \nabla_m \nabla_n \right)
    \Phi \, ,
\eeq
from which we see $m_{KK} \sim 1/r^2$.

\ssubsubsection{Light Fermions}
A similar calculation can be made for the spectrum of light
fermions, and leads to the following light fermion
spectrum:
\vspace{3mm}
\begin{center}
\begin{tabular}{ccc}
No. & Spin & Field \\
1 & 3/2 & $\psi_{\mu}(x)$ \\
6 & 1/2 & $\chi(x)$, $\lambda(x)$, 4 combinations of
$\psi_{m}(x,y)$
\end{tabular}
\end{center}
\vspace{3mm}

For our purposes it is fruitful to determine how these fields
assemble into multiplets of the unbroken 4D supersymmetry. The
identification of these multiplets may be explicitly obtained by
using the supersymmetry transformations of eq.~\pref{E:susy} --
such as by following the arguments of ref.~\cite{BFQ} -- and leads
to the following:
\begin{itemize}
\item
The massless gravitino required by the unbroken supersymmetry is
the partner of the graviton.
\item
Three massless gauginos arise as partners of the $SO(3)$ gauge
bosons. These fermions come from the higher-dimensional gravitino
due to the simultaneous existence of a Killing spinor and three
Killing vectors.
\item
A massless fermion combines with $s$ and $a$ into a massless
chiral multiplet, whose complex scalar part may be written $S =
\frac12( s + ia)$.
\item
Two fermions, with masses $m^2 \sim g^2/s$, join $t$ and $A_\mu +
\partial_\mu b/f$ to fill out a massive spin-1 multiplet. This massive
multiplet can be regarded as the result of a massless spin-1
multiplet `eating' the chiral multiplet whose complex scalar part
is $T = \frac12(t + ib)$ {\it via} the Higgs mechanism.
\end{itemize}

Once the fermions are chosen to transform in the standard way
under $N=1$ 4D supersymmetry, they do not carry the $U(1)$ gauge
charge, even though the 6D fermions did -- {\it c.f.}
eq.~\pref{E:covderiv}. In detail this happens because the 4D
supersymmetry eigenstates are related to the 6D fermions by powers
of the scalar $e^{ib}$, which cancel the 6D fermions'
transformation properties. Only the gauginos of the low-energy
theory transform nontrivially under the $SO(3)$ gauge symmetry.

\section{The 4D Effective Theory}
\label{S:4DETI}
Since the low-energy theory has an unbroken $N=1$ supersymmetry,
it must be possible to write it in the standard $N=1$ supergravity
form. From the previous section we see that the matter superfields
in terms of which the action below the compactification scale is
expressed are the massless chiral multiplet, $S$ and the three
massless gauge multiplets, $A_a, a = 1,2,3$.

Since our interest is in exploring the shape of the scalar
potential as a function of both $r$ and $\phi$, it is useful to
extend the effective action to also include the massive chiral
field, $T$, and the massive $U(1)$ gauge multiplet, $A$, which is
related to it by the Higgs mechanism. The effective theory
obtained in this way is not a {\it bona fide} Wilsonian action
when evaluated along the flat direction, however. It is not
because the mass of these fields are $m_t^2 \sim g^2 \,
e^\phi/r^2$ and $m_A^2 \sim e^{-\phi}/(g^2 r^6)$, which are the
same order of magnitude as the generic Kaluza-Klein mass,
$m_{KK}^2 \sim 1/r^4$, when evaluated along the trough of the
potential (along which $g^2 r^2 e^\phi \sim O(1)$).\footnote{We
thank G.~Gibbons and C.~Pope for correcting an error concerning
the relative sizes of $m_{KK}$ and $m_t, m_A$ in the original
version of this paper.} We may nevertheless choose to `integrate
in' these modes in the spirits of
refs.~\cite{GCondensation,Wilsonvs1PI}, with the idea that
parametrically their masses depend differently on $r$ and
$e^\phi$, and so there can be regions of field space away from the
bottom of the potential's trough for which they are systematically
light compared to $m_{KK}$.

In order to completely specify all of the terms of the 4D
supergravity action, it suffices to identify the K\"ahler
function, $K(S,S^*,T,T^*,A,A_a)$, the gauge kinetic functions,
$H_3(S,T)$ and $H_1(S,T)$ for the $SO(3)$ and $U(1)$ gauge groups,
the superpotential, $W(S,T)$, and the Fayet-Iliopoulos term, $\xi$
\cite{Cremmer,BW}.

\subsection{The Lowest-Order Action}
In this section we determine these functions classically, by
comparison with the direct truncation of the 6D
action~\cite{Witten,BFQ}, followed by a discussion of the kinds of
corrections which may be expected for the result \cite{BFQ,SNRT}.

\ssubsubsection{The K\"ahler Function}
An important constraint on $K$ arises because $b$ is eaten by the
$U(1)$ gauge field, $A_\mu$, since this implies its derivatives
can only enter ${\cal L}$ through the gauge-invariant combination
$\partial_\mu b + f A_\mu$. One infers from this that the
superfields $T$ and $A$ must enter the K\"ahler function only
through the combination $T + T^* + c A$, for a real constant $c$
to be determined below. Similarly, the shift symmetry $a \to a +$
(constant) implies $K$ can depend on $S$ only through the
combination $S + S^*$.

The form for $K$ is most easily read off from the scalar kinetic
terms, which in the Einstein frame must take the form ${\cal
L}_{\rm kin} = - K_{ij^*}
\partial_\mu z^i \partial^\mu z^{j*}$ for generic complex scalar
fields, $z^i$. (As usual subscripts here denote derivatives of $K$
with respect to the relevant scalar field, evaluated with all
fields except the scalars vanishing.) Comparing this with the
direct truncation calculation of the kinetic terms for $r$ and
$\phi$, eq.~\pref{E:rphikin}, gives the result
\eq \label{E:Kahlerc}
    K_{tr} = - \log\Bigl( S+ S^* \Bigr) - \log \Bigl( T + T^* +
    c A \Bigr).
\eeq

\ssubsubsection{The Gauge Kinetic Functions}
The gauge kinetic functions are more constrained than is the
K\"ahler function since they must depend holomorphically on their
arguments. They may be read off from the gauge boson kinetic
terms, which must have the general form ${\cal L}_{\rm kin} = - \;
\frac14 \,\Bigl[ (\hbox{Re} \; H_1) \, F_{\mu\nu} F^{\mu\nu} +
(\hbox{Re} \; H_{3}) \, F^a_{\mu\nu} F_a^{\mu\nu} \Bigr]$.
Alternatively, for some purposes they may be more simply obtained
from the related terms ${\cal L}_{\theta} = - \; \frac14 \,\Bigl[
(\hbox{Im} \; H_1) \, F_{\mu\nu} \tilde F^{\mu\nu} + (\hbox{Im} \;
H_{3}) \, F^a_{\mu\nu} \tilde F_a^{\mu\nu} \Bigr]$, since the
imaginary parts of $S$ and $T$ appear in more restricted ways in
the reduction of the 6D action.

Comparing these with the direct truncation of the 6D action gives
the result: Re$H_1 = e^{-\phi} \, r^2 = s = 2 \, \hbox{Re} \, S$,
from which we find $H_1 = 2S$. (That the leading contribution to
$H_1$ must be proportional to $S$ follows from the recognition
that $H_1$ scales like $H_1 \to \sigma^2 \, H_1$ under the
classical transformation, eq.~\pref{E:ttransf}, together with the
transformations $S \to \sigma^2 \, S$ and $T \to T$.) The
higher-derivative corrections of eq.~\pref{E:GkinCorr} which
follow from anomaly cancellation correct this result to give
\eq \label{E:H1ST}
    H_1 = 2 \left( S + {\tilde v \over v} \, T \right).
\eeq

A similar direct dimensional reduction for the $SO(3)$ gauge
fields is more involved, since the massless mode is a linear
combination of the fields $A_\mu(x,y)$ and $g_{\mu n}(x,y)$
\cite{RDSS}. (The necessity for mixing between $A_\mu$ and $g_{\mu
n}$ may be seen by performing a local $SO(3)$ transformation
corresponding to the general coordinate transformation $y^m \to
\xi^m(x,y) = \omega^a(x) \, K^m_a(y)$, where $K^m_a(y), a = 1,2,3$
are the three Killing vectors which generate the $SO(3)$
isometries of the sphere. Under this transformation the massless
4D gauge potential must transform as $\delta A^a_\mu =
\partial_\mu \, \omega^a + \cdots$.) Consequently, the $SO(3)$
gauge kinetic function acquires contributions from both the 6D
Einstein-Hilbert and Maxwell terms of the action.

For our purposes the details of this reduction are not necessary
in order to conclude that $H_3$ is given by an expression very
much like eq.~\pref{E:H1ST}:
\eq \label{E:HST} H_3 = 2(\alpha S +  \beta T), \eeq
for constants $\alpha$ and $\beta$ which are given in terms of the
anomaly coefficients $k, v$ and $\tilde v$. This conclusion is
most easily established by considering ${\cal L}_\theta$ and
recognizing that for the $SO(3)$ fields these terms are linear in
$a = 2 \, \hbox{Im}\, S$ and $b = 2\, \hbox{Im} \, T$. This is
most easily seen from the contribution of the Lorentz Chern-Simons
term in $G_{MNP}\, G^{MNP}$ and from the Green-Schwarz anomaly
cancelling term, eq.~\pref{E:AnomCan}. Linearity in $a$ is as
expected from the scaling property, eq.~\pref{E:ttransf}, together
with the transformation properties of $S$ and $T$.

\ssubsubsection{The Scalar Potential}
The constant $c$ in the K\"ahler potential, the superpotential,
$W$, and the Fayet Iliopoulos term, $\xi$, are fixed by
considering the scalar potential, eq.~\pref{E:stpot}. This must
agree with the general supergravity form $V = V_D + V_F$, where
\eqa \label{E:SGpot}
    V_F &=& e^K \; \Bigl[ (K^{-1})^{ij^*} \left( W_{i} + K_{i} W
    \right) \left( W_{j} + K_{j} W \right)^* - 3 |W|^2 \Bigr],
    \nn\\
    V_D &=& - \, \frac12 \; \left(\hbox{Re} \; H_1\right) D^2 -
    \, \frac12 \; \left( \hbox{Re} \; H_3 \right)D_a D_a
    \, , \eeqa
where $D_a$ and $D$ are the auxiliary fields for the two factors
of the gauge group, which we have not yet integrated out (hence
the potential's unusual sign). Here, as usual, $(K^{-1})^{ij*}$
denotes the inverse of the matrix of second derivatives,
$K_{ij*}$.

Given that neither $S$ nor $T$ carry $SO(3)$ gauge quantum
numbers, we see that $D_a = 0$ must be used when comparing with
the truncated 6D action. Since $T$ does transform under $U(1)$,
$D$ can be nonzero and, from the K\"ahler and gauge kinetic
functions found above, the $U(1)$ $D$ terms of the low-energy
action arise from the following terms:
\eqa
    {\cal L}_D &=&  s \, \left( - \; \frac14 \, F_{\mu\nu}
    F^{\mu\nu} + \frac12 \, D^2 \right) + D \left( \xi  +  \left.
    {\partial K \over \partial A} \right|_{A=0} \right)  \nn\\
    &=& s \, \left(- \;  \frac14 \, F_{\mu\nu} F^{\mu\nu} +
    \frac12 \, D^2 \right) + D \left( \xi - { c \over T + T^*}
    \right) \, . \eeqa
Here $\xi$ is the Fayet-Iliopoulos term, which is permitted only
for $U(1)$ gauge fields. Notice that consistency requires we use
only the lowest-order expression $H_1 = 2 S$ when comparing with
the action given above.

Integrating out $D$ implies the saddle-point condition
\eq D = -{1 \over s} \; \left( \xi - {c \over T+T^*} \right) \, ,
\eeq
and so leads to the potential
\eq V_D = + {1 \over 2\, s} \; \left(\xi - {c \over t} \right)^2
\, . \eeq
Comparing this with eq.~\pref{E:stpot} we read off:
\eq \xi = \pm 2 \, g \qquad \hbox{and} \qquad c = \pm \; {1 \over
2\, g} \, . \eeq
This appearance of Fayet-Iliopoulos terms when the fermion content
has a gauge anomaly is discussed in more general terms in
ref.~\cite{FITerms}.

Since this completely accounts for the scalar potential and
supersymmetry is unbroken, we conclude that the superpotential
vanishes:
\eq \label{E:Wis0} W = 0. \eeq

\subsection{Perturbative Corrections}
The above expressions for $K, W, H_1, H_3$ and $\xi$ are derived
by classically truncating 6D supergravity, and so in principle
they only apply strictly in the limit that $r \to \infty$ and
$e^\phi \to 0$, since it is only in this limit that the
corrections to truncation vanish. In this way we see that the
truncation results are approximations to the full expressions
which work for the region $s,t \to \infty$ of the space of moduli.

For sufficiently large $s$ and $t$ -- both of which are large if
$1/r^2 \ll e^\phi \ll 1$ -- the corrections to the truncation may be
computed order-by-order in a low-energy, weak-coupling expansion
in powers of $1/s$ and $1/t$. (Some of these corrections have
already been computed for the gauge kinetic functions above.)
Fortunately, the interplay of 6D and 4D supersymmetry strongly
restricts the form which such corrections may take
\cite{BFQ,SNRT}. As usual, these implications are stronger for the
holomorphic functions $H_1, H_3$ and $W$ than they are for $K$ and
so we discuss these two cases separately.

\ssubsubsection{Holomorphic Functions}
We first discuss the form which perturbative corrections may take
for the holomorphic functions of the supergravity action. For the
superpotential, $W$, as has been known for a long time \cite{NRT},
holomorphy completely forbids perturbative corrections from
arising within perturbation theory \cite{NRT}, leading to the
complete absence of correction to $W$ to all orders in $1/T$ and
$1/S$ \cite{BFQ,SNRT}. This leaves eq.~\pref{E:Wis0} as the
complete prediction to all orders.

Perturbative corrections to $H_1$ and $H_3$ do arise at one loop,
and are given by the $S$-independent terms in eqs.~\pref{E:H1ST}
and \pref{E:HST}. No further corrections beyond these are allowed
to all orders in perturbation theory, however. This may be seen
from the symmetry under shifts in Im$\,S$, which is broken only by
the Chern-Simons terms in the field strength for $B_{MN}$, since
these determine the gauge transformation properties of $B_{MN}$,
eq.~\pref{E:Btransf}. As we have seen, these Chern-Simons terms
are themselves related to the Green-Schwarz action which cancels
the gauge anomaly of the 6D fermions, and this connection with the
anomaly precludes there being additional terms of this form which
are generated beyond one loop. We see that expressions
\pref{E:H1ST} and \pref{E:HST} are therefore the complete
predictions -- up to the additions of $S$- and $T$-independent
constants -- for $H_1$ and $H_3$ to all orders in $1/S$ and
$1/T$.\footnote{As has been noted elsewhere, since this argument
relies on holomorphy it strictly applies only to the Wilson
action, and not necessarily to the generator of 1PI vertices
\cite{Wilsonvs1PI, GCondensation}.}

Perturbative corrections are necessarily concentrated into the
K\"ahler function, and it is to a discussion of these that we now
turn. These come in two forms.

\ssubsubsection{K\"aher Function: Powers of $1/t$}
Corrections to the K\"ahler function can arise, and do so
independently as powers of $1/t$ and $1/s$, since these have roots
in the more microscopic theory as independent expansions in powers
of $e^\phi$ and $1/r$. This may be seen explicitly by considering
two types of corrections to the lowest-order action in the 6D
theory, as we now do. We start with powers of $1/t$, which play an
important role in what follows, and which we now argue correspond
to the contributions under dimensional reduction of
higher-derivative corrections to the 6D effective theory.

The simplest way to identify corrections to $K$ is to compute the
corrections to the K\"ahler metric by examining the kinetic terms
of the scalars $r$ and $\phi$ in the 4D effective action. Examples
of higher-derivative corrections to these kinetic terms are the
contributions of higher-curvature terms to the radion kinetic
energy. For instance a higher-curvature correction to the
Einstein-Hilbert action (in the string frame) in six dimensions
\eq
    {{\cal L}_{SF} \over e_6} \sim e^{-2 \phi} \Bigl[R_s + k_n R_s^n \Bigr]
\eeq
becomes, in the Einstein frame
\eq
    {{\cal L}_{EF} \over e_6} \sim R + k_n e^{-(n -1)\phi} R^n ,
\eeq
due to the rescaling $g_{MN} \to e^\phi g_{MN}$ which is required
to remove $\phi$ from in front of the Einstein-Hilbert part of the
action. On dimensionally reducing we extract one factor of the 4D
Ricci tensor, $R_{\mu\nu} \propto \partial_\mu r \,
\partial_\nu r/r^2$, from $R^n$, with the remaining factors being
proportional to the two-dimensional curvature: $R_{(2)}^{n-1}
\propto (1/r^2)^{n-1}$, leading to the 4D kinetic term
\eq
    {{\cal L}_{\rm kin} \over e_4} \sim {\partial_\mu r \,
    \partial^\mu r \over r^2}\left[ 1 + {k_n \over (e^\phi
    r^2)^{n-1}} \right] \sim {\partial_\mu r \,
    \partial^\mu r \over r^2}\left[ 1 +  {k_n \over t^{n-1}}
    \right].
\eeq
For instance, in string theory such corrections could arise from
sigma-model corrections at string tree level.


\FIGURE{ \epsfig{file=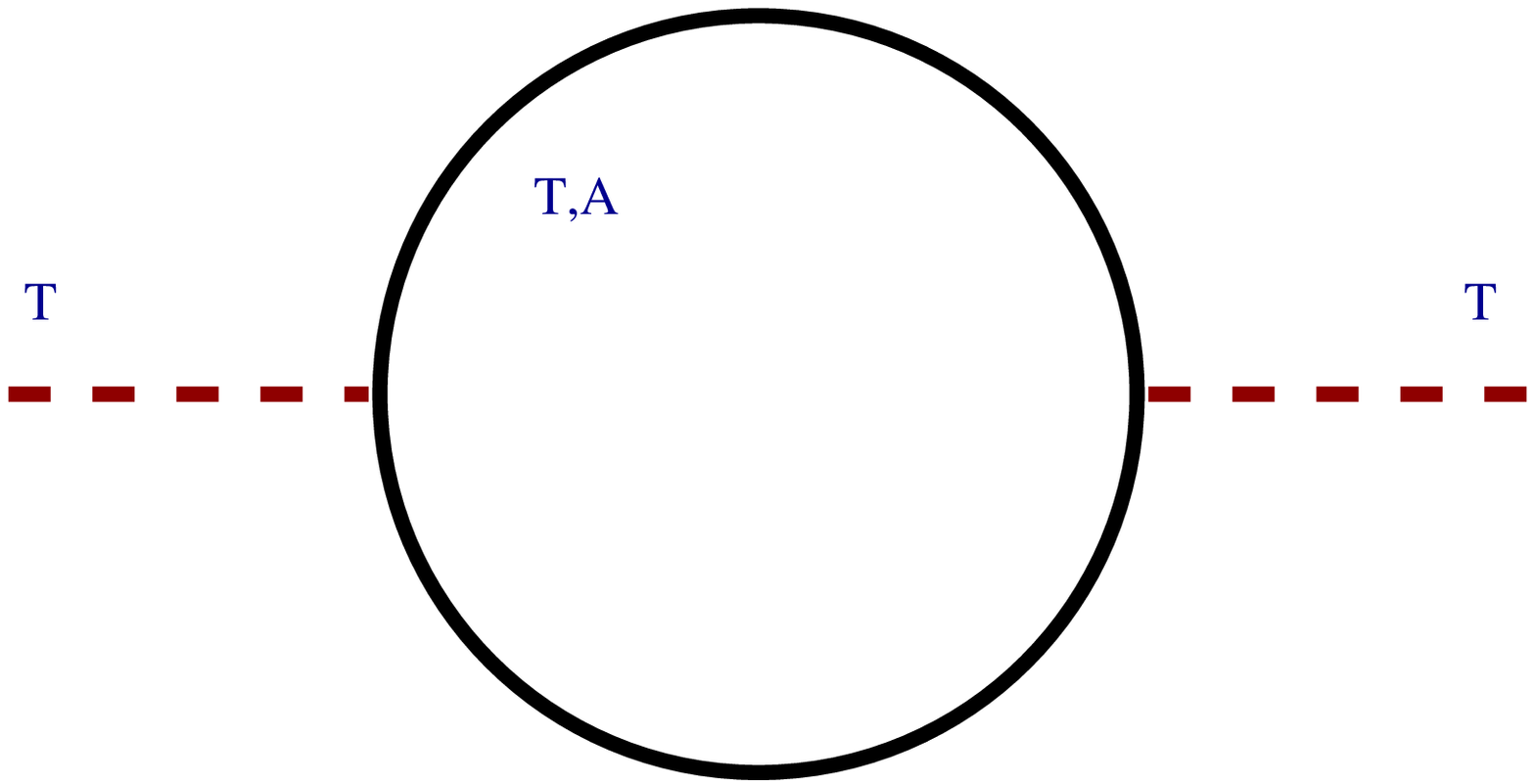, width=3.5in}
    \caption[Figure 1]{One-loop diagram contributing
to the kinetic term of the field $T$. The internal lines are also
$T$ and/or the $U(1)$ gauge field.\label{figure1}}}

\ssubsubsection{K\"aher Function: Powers of $1/s$}
Powers of $1/s$ can arise due to loops within the 4D theory
itself. For instance, consider the one-loop correction to the
kinetic term for $T$ which is induced by the graph of
Fig.~(\ref{figure1}). Since the 4D loop integrals diverge
(quadratically) in the ultraviolet, they are insensitive to the
masses of the multiplets in the loop and we can ignore the mixing
between the $T$ multiplet and the $U(1)$ gauge multiplet. If we
take all internal lines in Fig.~(1) to be $T$ fields, then the
vertices are of order $\partial^3 K /\partial t^3 \sim 1/t^3$,
while each propagator contributes $[K_{TT^*}(p^2 + m_t^2)]^{-1}$
with $K_{TT^*} \sim 1/t^2$. Taking the 4D ultra-violet cutoff to
be $m_{KK}^2 \sim 1/r^4 \sim M_p^2/({st})$, we estimate:
\eq \delta K_{TT^*}(T \; \hbox{loop}) \sim {m_{KK}^2 \over (4
\pi)^2 M_p^2} \left( {1\over t^3} \right)^2 \; \left( {t^2}
\right)^2 \sim {1 \over (4 \pi)^2 \, s \, t^3} . \eeq

Alternatively, if one of the internal lines of Fig.~(1) is a
$U(1)$ gauge multiplet, then the coupling between $V$ and $T$,
given by the lowest order K\"ahler function $K = - \log (T+T^* +
cV)$, is of order $\partial^3 K/\partial V \partial T
\partial T^* \sim c/t^3$, and we see that each vertex of Fig. (1)
contributes a factor $c/t^3$. Taking $s \gg t$ the gauge
propagator contributes a factor of $1/s$, so we estimate:
\eq \delta K_{TT^*}(T-V \; \hbox{loop}) \sim {m_{KK}^2 \over (4
\pi)^2 M_p^2} \left( {c\over t^3} \right)^2 \; \left( {t^2 \over
s} \right) \sim {1 \over (4 \pi)^2 \, s^2 \, t^4} , \eeq
where we use the lowest-order condition $t\sim 1/g^2 \sim c^2$.

\ssubsubsection{K\"aher Function: Logarithms of $1/s$}
Having seen how powers of $1/t$ and $1/s$ control the
modifications to $K$, we next consider the possibility that more
subtle types of corrections arise, which depend logarithmically on
$1/s$. Indeed, logarithmic dependence on coupling constants is
known to arise in 4D physics if the energies of some low-lying
states, $E_l$, are suppressed by powers of coupling constants,
$g$, relative to higher-energy states, $E_h$: $E_l \propto g^n
E_h$. In this case logarithms of couplings can arise as logarithms
of energy ratios: $\log(E_h/E_l) \sim n\, \log(1/g)$. The most
well-known example of this type is perhaps the QED prediction for
the Lamb shift, which involves a famous factor of $\log(1/\alpha)$
\cite{WbgVolII}. The potential for these kinds of logarithms
exists in the low-energy 4D theory arising from the 6D
supergravity compactification considered above because of the
existence of hierarchies of mass scales. For instance if string
states have masses $M_s \sim M_p$ then these are very different
from typical compactification scales, $m^2_{KK} \sim 1/r^4 \sim
M_p^2/(s\, t)$, which are suppressed by powers of the small
quantity $1/(s\, t)$. If logarithms of such ratios arise they can
give rise to logarithms of $s$ and $t$, of the form
\eq
    \log\left( \frac{M_p^2}{m_{KK}^2} \right) \sim \log ( s\, t ) ,
    \qquad \log\left( \frac{M_p^2}{m_A^2} \right) \sim \log(g^2
    s \, t^2) , \qquad \log \left( \frac{M_p^2}{m_t^2} \right) \sim
    \log(s/g^2) \, ,
\eeq
all of which are similar in size when evaluated along the bottom
of the scalar potential.

Some of these logarithms can arise in the four dimensional theory
due to the appearance of large logarithms in the running of the
couplings, and if so their appearance can be understood (and often
re-summed) using standard renormalization-group arguments
\cite{WbgRG,BM}. To this end imagine running the 4D effective
theory within the 4D theory, where the running of the inverse
couplings, $H_1(S,T)$ and $H_3(S,T)$, is given by
\eqa \label{E:HRGeqs} \left.H_1(S,T)\right|_\mu &=&
\left.H_1(S,T)\right|_{\mu_0}
    + b_1 \, \log\left({\mu^2 \over {\mu_0}^2} \right) \cr
\left.H_3(S,T)\right|_\mu &=& \left.H_3(S,T)\right|_{\mu_0}
    + b_3 \, \log\left({\mu^2 \over {\mu_0}^2} \right) \, ,
\eeqa
with $b_1$ and $b_3$ are the standard supersymmetric one--loop
beta--function coefficients for the $U(1)$ and $SO(3)$ gauge
groups, respectively. Using the expressions \pref{E:H1ST} and
\pref{E:HST}, we may solve for the running of $s$ and $t$
\eq \label{E:RGeqs}
    s(\mu^2) = s_0 + b_s \, \log\left( {\mu^2\over \mu^2_0}
\right), \qquad \hbox{and} \qquad t(\mu^2) = t_0 + b_t \,
\log\left( {\mu^2 \over \mu^2_0} \right), \eeq
where, as long as the equations are non-singular, $b_s$ and $b_t$
are linear combinations of $b_1$ and $b_3$ depending on the
coefficients of $S$ and $T$ in $H_1$ and $H_3$.

The dependence of large logarithms on masses, $m$, smaller than
$m_{KK}$ may then be traced by running these 4D couplings down to
$\mu = m$ from $\mu_0 = m_{\scriptscriptstyle KK}$. This gives
logarithms of the form $\log(m_{KK}^2/m^2)$, such as
\eq \log \left( {m_{\scriptscriptstyle KK}^2 \over m^2_A } \right)
\sim \log \left( { g^2} t  \right) + \hbox{constant} \, ,\eeq
which are large if $t \gg 1/g^2$ (away from the bottom of the
potential).

We see from these considerations that the existence of logarithms
of $s$ and $t$ in the corrections to $K$ are not unlikely. Without
performing a more sophisticated calculation it is difficult to pin
down the precise power of $s$ and/or $t$ which appears inside the
logarithm. This is because low-energy logarithms like
$\log(m_{\scriptscriptstyle KK}^2/m_A^2)$ can in principle combine
with other large logarithms which arise purely from the
high-energy theory, such as $\log(M_p^2/m_{\scriptscriptstyle
KK}^2)$ to give new logarithms like $\log(M_p^2/m_A^2)$. We
therefore parameterize this possibility by writing the resulting
full RG-improved K\"ahler function as
\eq
\label{E:Krun}
K = -\log\left[s - b_s \log\left(s t^a\right) + k_s \right]
    - \log\left[t-b_t\log\left(s t^a\right) + k_t + cA \right],
\eeq
where $a$, $k_s$ and $k_t$ are order-unity constants.

Notice that expanding eq.~\pref{E:Krun} in powers of $1/s$ and
$1/t$ gives the corrections to $K$ to leading order in $1/t$ and
$1/s$, but to all orders in $(1/s)\log(st^a)$ and/or $(1/t)
\log(st^a)$. This observation will become important later when we
find minima for the scalar potential.

\subsection{Nonperturbative Effects in 4D}
Given the above semiclassical approximation to the functions $K$,
$H$, $H_{ab}$ and $W$, we may use general knowledge of 4D $N=1$
supersymmetric theories to understand the physics at energies much
below the compactification scale. In particular, our interest is
in the existence of any other mass scales at very low energies
which might lift the degeneracy of the flat direction described by
$S$.

It is useful to re-instate the Planck mass and to identify the
mass scales which arise in the low-energy 4D supergravity. These
are:
\begin{itemize}
\item
The 4D Planck mass: $M_p^2 = 1/\kappa_4^2 = 1/(8\pi G_4)$, as
defined by the 4D graviton couplings.
\item
The 4D cutoff: $m_{KK} \sim 1/r^2 \sim M_p/(st)^{1/2}$, which
defines the scale above which the theory is no longer efficiently
described by a 4D lagrangian.
\end{itemize}

To these semiclassical mass scales should be added a new,
nonperturbative one: $\Lambda \sim \mu \, \exp \left[- (\nu s(\mu)
+ \lambda \, t(\mu))/3 \right]$, where $\nu$ and $\lambda$ are
positive constants which are related to the renormalization-group
coefficients for $s$ and $t$ by the condition that $\Lambda$ be
independent of renormalization point $\mu$. Using
eqs.~\pref{E:RGeqs} this implies
\eq
\label{E:Beqs}
\nu \, b_s + \lambda \, b_t = \frac32 \, . \eeq

This new scale arises because the low-energy theory's $SO(3)$
gauge theory is asymptotically free, with $\Lambda$ defining the
confinement scale where its effective coupling becomes strong. At
this scale the gauginos of the $SO(3)$ theory condense
\cite{GCondensation}, and because of this condensation (together
with the absence of matter fields carrying approximate global
chiral symmetries) the $SO(3)$ gauge sector acquires a gap in its
spectrum which is of order $\Lambda$. The massive energy
eigenstates which result are the $SO(3)$-singlet bound states of
the gluons and gluinos.

As is well known, this condensation dynamically generates a
superpotential in the low-energy theory \cite{GC,GCondensation},
which is of order $\Lambda^3$:
\eq \label{E:npSP}
    W  = w_0 \; \exp[ - \nu \, S - \lambda \,
    T ] , \eeq
for some constant $w_0 \sim \mu^3$.

This superpotential contributes to the scalar potential for $s$
and $t$ by generating a nonzero $V_F$, which was absent
semiclassically. It is this new term which is responsible for the
qualitatively new features of the low-energy theory: the lifting
of the flat direction for $s$.

\section{Dynamics of the Flat Directions}
\label{S:FlatDirections}
We have seen that the strongly-coupled $SO(3)$ gauge couplings
dynamically generate a superpotential at low energies, and so the
two terms, $V_D$ and $V_F$ conflict in what they would like the
fields $t$ and $s$ to do. The semiclassical term, $V_D$, is
minimized when $t \sim 1/g^2$, while the nonperturbative term,
$V_F$, is minimized when $t \to \infty$. These cannot be
simultaneously minimized and so a compromise must be struck for
which at least one of $V_F$ or $V_D$ is nonzero. We find that the
vacuum to which this competition between $V_F$ and $V_D$ leads
depends in a crucial way on the form of the corrections to $K$
discussed above.

\subsection{Dilaton Runaway}
As a first approximation to the shape of this potential, we
consider the superpotential, eq.~\pref{E:npSP}, but ignore all
corrections to the leading semiclassical K\"ahler function,
eq.~\pref{E:Kahlerc}, and gauge kinetic functions. This leads to a
scalar potential of the form $V = V_D + V_F$, where $V_D$ may be
read off from eq.~\pref{E:stpot}, and $V_F$ is given by:
\eq \label{E:VFdef}
    V_F(s,t) = {|w_0|^2 \over s t} \, e^{-\nu \, s - \lambda \, t}
    \; \left[ \left( 1 + \nu \, s \right)^2 + \left( 1 + \lambda t
    \right)^2 - 3 \right] . \eeq

If $\Lambda \ll M_p$ we have $w_0 \ll 1$ and the minimum for $t$
is close to the zero of $V_D$: $1/t = 4g^2 + O(|w_0|^2)$. To
linear order in $|w_0|^2$ the potential for $s$ then becomes
$V_{\rm eff}(s) \approx V_F(1/t=4g^2)$, and so
\eq
      V_{\rm eff}\left(s  \right)  =
      {4 g^2 \, |w_0|^2 \over s} \, e^{-\nu \, s - \lambda /( 4 g^2)}
    \; \left[ \left( 1 + \nu \, s \right)^2 + \left( 1 +
    {\lambda\over 4 g^2} \right)^2 - 3 \right] .\eeq
We find that the potential for $s$ which is generated in this
approximation does not have any minima for positive $s$ besides
the runaway solution for which $s \to \infty$. This is the
familiar dilaton runaway, with the $SO(3)$ gauge coupling
generically driven to zero as $s$ runs off to infinity.

\subsection{Dilaton Stabilization}
The weak part of the previous analysis is the use of the
lowest-order K\"ahler function, eq.~\pref{E:Kahlerc}, despite
using a nonperturbative expression for the superpotential. We now
show that using the renormalization-group-improved expression,
eq.~\pref{E:Krun}, can generate a potential for $s$ which can have
other minima besides the dilaton runaway.

We begin with the K\"ahler function,
\eqa \label{E:Kform}
    K(s,t) &=&
- \log \left[ s + {b_s \over 2} \, \log \left( {s t^a \over q} \right)
    \right] - \log \left[ t
    + {b_t \over 2} \, \log \left( {s t^a \over q} \right)
    \right] \cr
&& \qquad
+\; O\left({\log (st^a/q) \over s^2},{\log (s t^a/q) \over t^2}\right), \cr
&\equiv&
-\log s_0 - \log t_0 + O\left({\log  \over s^2},{\log  \over t^2}\right),
 \eeqa
where $a$ and $q$ are constants, and where $s_0, t_0$ are the
fields evaluated at the high scale.  Notice that changing from
$s_0, t_0$ to $s, t$ in $K$ (and then constructing the scalar
potential $V$) is {\em not} simply the same as performing a
trivial change of variables on the potential $V$ itself. It is not,
because this change is not a holomorphic redefinition of $S$ and
$T$.

\def\kahler{K\"ahler}
\def\detK{||K||}

Under the assumption that $|w_0| \ll 1$ we may compute the
effective potential as before, by first minimizing $V_D(s,t)$ to
obtain $t = t(s)$ and then examining $V_{\rm eff}(s) \approx
V_F[s,t(s)]$. The minimum of $V_D$ occurs when
\eq
\label{E:Dcond}
K_T + \epsilon = 0 ,
\eeq
where $\epsilon$ is a constant which is of order $g^2$ and
\eq
    K_T = {\partial K \over \partial T}
      \approx - \; {1 \over t_0}
    - {a\over 2}\,\left( \nth{t} {b_t \over t_0} + \nth{s} {b_s \over s_0}
    \right)
\eeq
In the case where $a=0$ \pref{E:Dcond} can be solved analytically, so that
$V_D$ is minimized for $t(s)$ satisfying
\eq \label{E:tofs}
    t_0 \equiv
t + \frac12 \, b_t \, \log(s/q) \approx {1 \over \epsilon}  \, . \eeq
Solving this for $t(s)$ and using the result in $V_F[s,t(s)]$
gives $V_{\rm eff}(s)$.
The computation of $V_F$ requires the
inverse matrix:
\eqa
 (K^{-1})^{SS^*} &=& {K_{TT^*} \over \detK}, \cr
 (K^{-1})^{TT^*} &=& {K_{SS^*} \over \detK}, \cr
 (K^{-1})^{ST^*} &=& (K^{-1})^{TS^*} = -{K_{ST^*} \over \detK} \, ,
\eeqa
where $\detK$ is the determinant of the matrix $K_{ij^*}$.
To this order in the \kahler\
function we may ignore the difference between
$s$ and $s_0$ and $t$ and $t_0$ after taking derivatives, so
that
\eqa
   && K_{TT^*} = \nth{t_0^2} \cr
   && K_{SS^*} = \nth{s_0^2}\,
    \left(1 +  {\beta_t}
    + {\beta_t}^2
    + {3\beta_s} + {\beta_s}^2
    \right) \cr
   && K_{ST^*} = {\beta_t \over s_0 t_0} \cr
   && \detK = \nth{s_0^2\,t_0^2}
    \left( 1+ {\beta_t}
    + {3 \beta_s}
    + {\beta_s}^2 \right)\; ,
\eeqa
where $\beta_s = \hf b_s/s_0$, $\beta_t = \hf b_t/t_0 \equiv \hf \epsilon b_t$.
We also have the K\"ahler derivatives
\eqa
    D_S W &=& W_S + K_S \; W \cr
    &=& -\left[\nu + {\sigma \over s_0}\right]\, W \cr
    D_T W &=& W_T + K_T \; W \cr
    &=& - \; (\lambda + \epsilon) \,
    W, \nn
\eeqa
%
where both results are evaluated at $t = t(s)$ and we have used~\pref{E:Beqs},
and
{$\sigma =   {2}\left({7 \over 8} + \beta_t +\beta_s\right)$}.

One finds in this way the expression for $V_{\rm eff}(s) =
V_F[s,t(s)]$:
\eq \label{E:VFform}
    V_{\rm eff}(s) = { |w_0|^2 \over s_0 t_0}
    \, N(s_0,t_0) \, e^{-2\nu s_0 - 2\lambda t_0} \, ,
\eeq
where
\def\bt{\beta_t}
\def\bs{\beta_s}
\def\eps{\epsilon}
\eq \label{E:Nform}
    N(s_0,t_0)  =
    {\left(\nu s_0 + \sigma\right)^2
    -2 B\,(\nu s_0 + \sigma)
    +C\,t_0^2\, (\lambda+\eps)^2
    \over 1+\bt + 3 \bs + \bs^2 }
    -3
\eeq
with
\eq
B = \bt t_0 (\lambda + \eps) \qquad {\rm and} \qquad
C = 1+\bt+\bt^2 + 3\bs + \bs^2 \,.
\eeq
%
%
Notice that this reduces to the previous runaway potential in the
limit $b_t \to 0, b_s \to 0$, as long as we also set $\sigma=1$
(the conditions \pref{E:Beqs} do not apply in this limit).
$\bs$ and $\sigma$ are both $s_0$-dependent quantities.


\FIGURE{ \epsfig{file=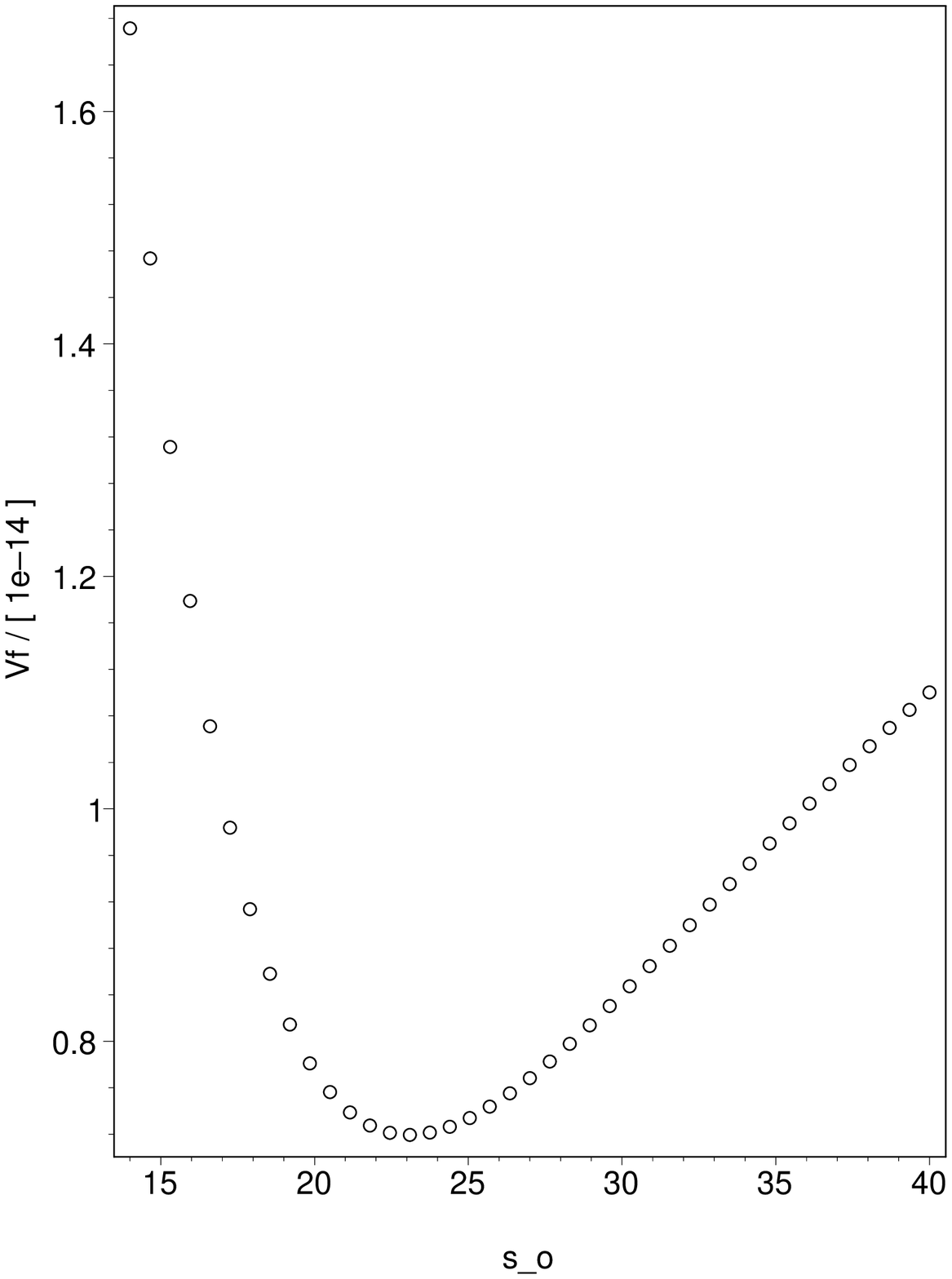, width=3.5in}
    \caption[Figure 1]{The effective potential for $s$
computed using the renormalization-group improved potential.
Parameters are chosen as described in the main
text.\label{figure2}}}

This potential is drawn in Fig.~\pref{figure2} for the choice, in
\pref{E:HST}, $\alpha = \beta = 1/800$, and with $\tilde{v}/v =
-9/4$, $q = 100$, $b_1 = -1/10$, $b_3 = 6/(4\pi)^2$, $\nu = 0.005$
and $\epsilon = .28$.  $\lambda$ is determined in terms of these
by the condition $\nu \, b_s + \lambda \, b_t = 3/2$, while $b_t$
and $b_s$ are determined from $b_1$ and $b_3$. For these choices
the potential is minimized by the field values $s = 34$ and $t =
8.6$, corresponding to $t_0 =1/\epsilon \approx 3.6$ and $s_0 =
23$.

It remains to be shown that values this small for $\alpha$ and
$\beta$ can be obtained from realistic string models. The above
discussion nonetheless suffices to make our main point that the
renormalization-group-improved K\"ahler function can produce
nontrivial minima for the modulus $s$. We have also found minima
having larger values of $\alpha$ and $\beta$
($|\alpha|,|\beta|\sim O(0.1)$), albeit for small values of $s$
and $t$ which lie at the limit of what can be understood
perturbatively in powers of $1/s$ and $1/t$. The generic existence
of such minima can be seen by observing that as long as the
potential has a maximum, and increases as one approaches the
origin ($s=t=0$) from the right, then a minimum must exist in
between (barring the existence of new singularities in this
regime). In the present case of the larger values of $\alpha$ and
$\beta$, maxima exist for $s$ and $t$ well within the perturbative
regime, allowing us to infer the existence of the minima at much
smaller field values: $s,t \sim O(0.1)$.

The quantities parameterizing the strength of the supersymmetry
breaking are given by the {\it vev} of the potential,
$\left.V_{\rm eff}(s)\right|_{\rm min}$, as well as the
expectation values of the auxiliary fields:
\eqa F_i &=& e^{K/2}D_i W, \cr M &=& e^{K/2} W/3 , \eeqa
where $i$
runs over $S$ and $T$. The mass of $S$ is approximately given by
\eq m_s^2 \approx  e^{-K} \left. {d^2 V_{\rm eff}(s) \over ds^2}
\right|_{\rm min}. \eeq
For the minimum described numerically above, we find $
\left<V\right> \approx 7 \times 10^{-15}$, $V'' \approx 9 \times
10^{-17}$, $F_s \approx -9 \times 10^{-8}$, $F_t  \approx -2
\times 10^{-6}$, $M \approx 2 \times 10^{-7}$ and $m_s^2  \approx
7\times 10^{-15}$, all in Planck units.  The
supersymmetry-breaking scale is therefore seen to be quite low,
compared to, say, the compactification scale $m_{KK} \sim
M_p/(st)^{1/2} \sim  M_p/10$.

\section{Discussion}
\label{S:Discussion}
We now summarize our results, and outline some of their potential
applications.

\subsection{Summary}
We have revisited the Salam-Sezgin compactification of gauged
$N=1$ 6D supergravity, and have computed the $N=1$ 4D supergravity
to which it leads at low energies. The low-energy field content to
which we are led is supergravity coupled to a supersymmetric $U(1)
\times SO(3)$ gauge theory plus several chiral multiplets which
describe the compactification's moduli. The low-energy theory has
the following properties:

\begin{itemize}
\item
The $U(1)$ multiplet `eats' one of the chiral multiplets {\it via}
the Higgs mechanism at the classical level giving these fields
masses which are comparable to the Kaluza-Klein mass scale,
$m_{KK} \sim \sim 1/r^2$, when evaluated along the trough at the
bottom of the classical scalar potential. This scalar potential
arises from the 4D point of view through a Fayet-Iliopoulos term
for the $U(1)$ gauge group.
\item
One combination of scalars parameterizes a flat direction which
remains massless to all orders in a semiclassical expansion, and
supersymmetry remains unbroken along this flat direction.
\item
The nonabelian $SO(3)$ gauge multiplet is asymptotically free and
once its coupling becomes large its gauginos condense and generate
a nonzero superpotential. The scalar potential which results from
this superpotential competes with the $U(1)$ $D$-term potential
and lifts the degeneracy of the flat direction.
\item
The vacuum to which the theory tends depends on the precise form
of the perturbative corrections to the K\"ahler potential. Using
the lowest-order result, $K = - \log s + \cdots$, leads to the
standard dilaton runaway, in which the massless field
parameterizing the flat direction runs off to infinity. In this
limit the $SO(3)$ gauge coupling vanishes and supersymmetry
remains unbroken.
\item
If we instead use the renormalization-group-improved version of
$K$, then the runaway can be stabilized for some choices of the
parameters. In this context the significance of the present
analysis is to identify a type of logarithmic dependence of $K$ on
$s$ which would be sufficient to stabilize the runaway. Of course,
we do not know yet whether the required parameters can actually
arise for low-energy perturbations about a real string vacuum.
However, we regard the potential rewards of their discovery to
provide sufficient motivation for taking a proper look.
\end{itemize}

At first sight, this last item appears to run contrary to a
standard argument by Dine and Seiberg against the possibility of
fixing the dilaton at weak string coupling \cite{DS}. This
argument essentially states that if the potential is a series in
$1/s$, then any minimum -- besides the runaway $s \to \infty$ --
must balance different terms in this series against one another,
and so be incalculable within the context of perturbation theory.

Despite this, we are able to find nontrivial minima in our
analysis for two reasons. First, given that $t$ and $s$ are large
at the minimum, but $(1/t)\log s$ is $O(1)$, we see that the
Dine-Seiberg argument is correct inasmuch as it states that the
potential is required to all orders in $(1/t)\log s$ in order to
determine its minima. Fortunately, this form is known by virtue of
the renormalization-group re-summation.

Second, one may ask how $t$ and $s$ could be large at the minimum
in the first place if there are no large parameters in the
potential. Although we have not exhaustively searched parameter
space for other solutions, it appears that we only obtain large
values for $s$ and $t$ when we choose small values for the
parameters $\alpha$ and $\beta$, and so this may explain the
origin of the nontrivial minima within perturbation theory. To the
extent that the appearance of these extra parameters which can be
tuned to get weak coupling are required, our analysis would be
similar to the older racetrack scenarios \cite{racetrack}.

In the end, it may be that realistic string models do not provide
$\alpha$ and $\beta$ of the required magnitude. We
regard the artifice of exploring the consequences of their being
small nonetheless to be of some value, because it allows us to
infer some evidence for the existence of minima in the
strong-coupled regime even in the cases where $\alpha$ and $\beta$
are larger. In this case the minima we find would be pushed into
the strong-coupling region for which our calculational methods do
not directly apply. Nevertheless the existence of these minima
still follows from the existence of a maximum for larger values of
$s$ and $t$, together with the general property that the potential
is positive and diverging as $s,t \to 0$. Indeed, the existence of
the maxima within the perturbative regime can be inferred for a
wider range of values for $\alpha$ and $\beta$ than can the
existence of the minima, leaving only the existence of the
singularity of the potential at small $s,t$ to be established
using more robust arguments.

\subsection{Potential Applications}
Both the runaway and stabilized scenarios have several potentially
interesting applications to low-energy phenomenology and to
cosmology, which we now briefly outline.

\ssubsubsection{Inflation}
The most conservative application of the above dynamics only
relies on the semiclassical stabilization of the $T$ modulus
through the $U(1)$ $D$-terms, and does not use the dilaton
stabilization mechanism. This application is to efforts to model
inflation within the brane-world \cite{BI,BI2}, and relies on the
recently-made observation \cite{BI2} that in some circumstances
supersymmetry can help make inflation easier to obtain if it is
being driven by brane physics.

A generic problem with obtaining exponential inflation from brane
physics arises because of the dynamics of the breathing mode of a
compactification. That is, even if an approximately-constant
potential energy can be found for a candidate inflaton which lives
on a particular brane, this potential energy tends not to lead to
de Sitter style inflation. It does not do so because a constant
energy on a brane is typically really constant only in the Jordan
frame tailored to the brane, and not in the Einstein frame for
which the Planck mass is constant. Instead, in the Einstein frame
it appears as a radius-dependent source of energy and so it causes
the sizes of the extra dimensions to evolve rather than causing
the ordinary 4 large dimensions to inflate.

Successful inflation along these lines therefore also requires the
extra dimensions to be stabilized, and it turns out that inflation
can be more easily obtained if it is a combination like $s =
r^2/e^\phi$ or $t = r^2 e^\phi$ which is fixed by the stabilizing
physics, rather than if $r$ or $\phi$ is separately fixed
\cite{BI2}. The semiclassical 6D model presented here is precisely
such a stabilization mechanism for $t$, and so suggests that a
similar mechanism might be employed to generate an inflationary
brane scenario.

Furthermore, our result of getting a positive cosmological
constant after fixing the moduli can be directly used as the
starting point of D-brane inflation along the lines of
references~\cite{BI}, where one imagines also adding an attractive
potential which causes the separation between D-branes to be the
inflaton field after all closed string moduli have been fixed.

\ssubsubsection{Supersymmetry Breaking}
If the runaway is stabilized the effective 4D model dynamically
breaks supersymmetry at a scale which can be naturally very small
compared with the compactification scale. If electroweak symmetry
breaking occurs at the supersymmetry-breaking scale, and if
fundamental scales like $M_s$ and $m_{KK}$ are chosen near the
Planck or GUT scales, then this would provide a Kaluza-Klein
realization for using dynamical supersymmetry breaking to
naturally generate the electroweak gauge hierarchy, along lines
initially proposed some time ago \cite{WittenHeirarchy}.

The low-energy implications of such a model may be inferred by
regarding the entire theory considered here to be the hidden
supersymmetry-breaking sector to which standard-model particles
are coupled \cite{SoftTerms}. As is easily verified, the large
hierarchy $s \gg t$ which the stabilization mechanism predicts
ensures that the auxiliary field for $S$ is the largest
supersymmetry-breaking {\it v.e.v.}. This makes the
phenomenological implications of this kind of supersymmetry
breaking the same as for a dilaton-dominated scenario. This has
the virtue of being among the most predictive kinds of
string-motivated supersymmetry-breaking scenarios, with definite
relations predicted for the spectrum of superpartners
\cite{SoftTerms}.

\subsection{String Theory Derivation?}

It is an interesting challenge to derive the 6D theory we started
with from string theory. First we note that typically 6D, $N=2$
supergravity is obtained from $K_3$ compactifications of type I or
heterotic strings. However the supergravity theory obtained in
this way is ungauged, whereas ours is gauged supergravity that
includes a nontrivial potential $\sim ge^{-\phi}$.

Recently it has been realized that massive 10D and gauged
supergravities in lower dimensions can be obtained either from
string compactifications on spheres \cite{chris}\, \footnote{We thank
 C. Pope for many discussions on these points.}
 or toroidal and related
compactifications, such as $K_3$, in the presence of RR or NS-NS
fluxes. In the latter case, typically a $p$-form $F_p$ integrated around
non-trivial cycles of the compact space $\Gamma_p$ can be
different from zero,
 $ \int_{\Gamma_p} F_p \neq 0$. The form $F_p$ can be expanded
in terms of harmonic forms $\omega^i_p$:
\eq
F_p\ =\ \nu^i \omega^i_p
\eeq
where the coefficients $\nu^i$ will correspond to the fluxes. In
particular these fluxes give rise to potentials precisely of the
form we started with, the flux being identified with the gauge
coupling constant of the effective gauged supergravity theory. In
this way several maximal gauged supergravity theories have been
obtained from toroidal compactifications with fluxes \cite{jan}.
Also compactifications on $K_3\times T_2$ have been considered
with fluxes in both $K_3$ and $T_2$. Furthermore, fluxes have also
been shown in type IIB and M theory to freeze geometric
moduli~\cite{sethi}, just as the $T$ field in the present model is
frozen.

We  have to recall  that the general $N=2$ 6D
supergravity has a more complicated spectrum of scalar fields than the
one we used, since the gauge group can be much larger than the
simplest $U(1)$ that we considered \cite{ergin}. In that case the potential is quadratic
in the hypermultiplet scalars, with overall factors of $e^{-\phi}$,
 a natural outcome of the fluxes in
string theory. It is not completely clear to us how to
derive precisely the Salam-Sezgin action from this class of
backgrounds yet, although it looks very suggestive.

Furthermore, string compactifications on spheres and related
geometries have also been successful in deriving gauged
supergravities. In particular the maximal 6D supergravity was
derived from an $S^4$ string compactification and a detailed
comparison of the potential was achieved with the $N=4$ gauged
supergravity of Romans \cite{romans}.\footnote{After completing
this work it was pointed out to us that fluxes over ${\cal
P}^1({\cal C})=S^2$ were considered in ref.~\cite{curio} in
Heterotic and Type II compactifications.} We leave as an open
question the possible derivation of our $N=2$ action from this
construction as well as any possible relationship with the $K_3$
backgrounds with fluxes. If such a construction is found then it
will be interesting since it will give rise to realistic string
compactifications on manifolds which are not Ricci flat (since at
least the 2-sphere is not), contrary to standard
beliefs.\footnote{After finishing this article we became aware of
earlier~\cite{sethi} and new~\cite{jan2} work presenting examples
of this type. It may be interesting to unravel any connection
between these constructions and our work.}

\subsection{Open Questions}
There are many questions left open at the moment.
Any application of the dilaton stabilization mechanism would be on
a much better foundation if the corrections of the form required
for $K$ could be computed as the consequence of an actual
microscopic underlying theory, such as from a viable string
configuration. We do not know how to obtain such a vacuum, but
regard the identification of the features required to enable
stabilization as a worthwhile first step towards identifying what
would be required of an underlying model in order to reproduce
this stabilization.

In general terms we see this model as a toy laboratory for
understanding more complicated string compactifications. It would
be interesting to have an explicit construction that precisely
generates this model, in which case it would be promoted to become
a consistent string vacuum. Nevertheless we believe that many of
the properties we discuss here apply more generally than just to
this particular model.

Another open question concerns the introduction of branes and
antibranes into the model, perhaps along the lines of \cite{sixd},
potentially leading to an implementation of a 6D brane-world
scenario. The concreteness of the construction has the virtue of
allowing the explicit study of their implications for
supersymmetry breaking and cosmology.

\acknowledgments We thank D. Grellscheid, A. Font, L. Ib\'a\~nez,
 C.
 Pope, R. Rabad\'an,  S. Theisen, P. Townsend and A. Uranga
 for interesting conversations. Y.A. and C.B.'s research
is partially funded by grants from N.S.E.R.C. of Canada and
F.C.A.R. of Qu\'ebec. S.P. and F.Q. are partially supported by
PPARC. F.Q. thanks the theory division at CERN for hospitality
during the conclusion of this work.



\begin{thebibliography}{99}

\bibitem{review} See for instance:
F. Quevedo
 hep-th/9603074 and references therein.

\bibitem{GC}
J.-P. Derendinger, L.E. Ib\'a\~nez and H.P. Nilles, {\it Phys. Lett.}
 {\bf B155} (1985) 65;
M. Dine, R. Rohm, N. Seiberg and E. Witten, {\it Phys. Lett.} {\bf B156} (1985) 55.

\bibitem{Flux}
S.~B.~Giddings, S.~Kachru and J.~Polchinski,
hep-th/0105097;
S.~Kachru, M.~B.~Schulz and S.~Trivedi,
hep-th/0201028.

\bibitem{Models}
G.~Aldazabal, L.~E.~Ib\'a\~nez and F.~Quevedo,
JHEP {\bf 0001} (2000) 031
[hep-th/9909172];
JHEP {\bf 0002} (2000) 015
[hep-ph/0001083];
G.~Aldazabal, L.~E.~Ib\'a\~nez, F.~Quevedo and A.~M.~Uranga,
JHEP {\bf 0008} (2000) 002
[hep-th/0005067];
R.~Blumenhagen, L.~Goerlich, B.~Kors and D.~L\"ust,
JHEP {\bf 0010} (2000) 006
[hep-th/0007024];
G.~Aldazabal, S.~Franco, L.~E.~Ib\'a\~nez, R.~Rabad\'an and A.~M.~Uranga,
JHEP {\bf 0102} (2001) 047
[hep-ph/0011132]; J.\ Math.\ Phys.\  {\bf 42} (2001) 3103
[hep-th/0011073].

\bibitem{BI}
G.~Dvali and S.~H.~H.~Tye, {\it Phys. Lett.}{\bf 450}{1999}{72}
[hep-ph/{9812483}];\\
C.P.~Burgess, D.~Nolte, M.~Majumdar, F.~Quevedo, G.~Rajesh and R.-J.~Zhang,
{\it JHEP} {07}({2001}){047} [hep-th/0105204];\\
G.~Dvali, S.~Solganik and Q.~Shafi, (unpublished)
[hep-th/0105203];\\
C.~Herdeiro, S.~Hirano and R.~Kallosh, {\it JHEP}
{0112}({2001}){027}
[hep-th/0110271];\\
J.~Garcia-Bellido, R.~Rabadan and F.~Zamora,
{\it JHEP} {0201}({2002}){036} [hep-th/0112147];\\
R.~Blumenhagen, B.~Korrs, D.~L\"ust and T.~Ott,
[hep-th/0202124]. For a recent review with many refrences see:
F. Quevedo, [hep-th/0210292].

\bibitem{BI2}
C.P.~Burgess, P.~Martineau, G.~Rajesh, F.~Quevedo and R.-J.~Zhang,
{\it JHEP}  {\bf 0203} (2002) 052, hep-th/0111025.

\bibitem{recent}
For recent discussions see:
M.~Dine and Y.~Shirman,
Phys.\ Rev.\ D {\bf 63} (2001) 046005
[hep-th/9906246];
S.~A.~Abel and G.~Servant,
Nucl.\ Phys.\ B {\bf 597} (2001) 3
[hep-th/0009089];
A.~Font, M.~Klein and F.~Quevedo,
Nucl.\ Phys.\ B {\bf 605} (2001) 319
[hep-th/0101186];
R.~Ciesielski and Z.~Lalak,
[hep-ph/0206134].

\bibitem{MS}
N. Marcus and J.H. Schwarz, {\it Phys. Lett.} {\bf 115B} (1982)
111.

\bibitem{NS}
H. Nishino and E. Sezgin, {\it Phys. Lett.} {\bf 144B} (1984) 187.

\bibitem{SS}
A. Salam and E. Sezgin, {\it Phys. Lett.} {\bf 147B} (1984) 47.

\bibitem{GandC}
S. Weinberg, {\it Gravitation and Cosmology}, Wiley, New York, 1972.

\bibitem{AGW}
L. Alvarez-Gaum\'e and E. Witten, {\it Nucl. Phys.} {\bf B234}
(1984) 269.

\bibitem{GSAC}
M.B. Green and J.H. Schwarz, {\it Phys. Lett.} {\bf B149} (1984)
117.

\bibitem{RDSSS}
S. Randjbar-Daemi, A. Salam, E. Sezgin and J. Strathdee, {\it
Phys. Lett.} {\bf B151} (1985) 351.

\bibitem{6DAC}
M.B. Green, J.H. Schwarz and P.C. West, {\it Nucl. Phys.} {\bf
B254} (1985) 327;\\
J. Erler, {\it J. Math. Phys.} {\bf 35} (1994) 1819
[hep-th/9304104].

\bibitem{6DSusy}
J.H. Schwarz, {\it Phys. Lett.} {\bf B371} (1996) 223
hep-th/9512953;\\
M. Berkooz, R.G. Leigh, J. Polchinski, J.H. Schwarz, N. Seiberg
and E. Witten, {\it Nucl. Phys.} {\bf B475} (1996) 115
hep-th/9605184;\\
N. Seiberg, {\it Phys. Lett.} {\bf B390} (1997) 169
[hep-th/9609161].

\bibitem{janber}
B.~de Wit and J.~Louis,
[hep-th/9801132].


\bibitem{Sagnotti}
A. Sagnotti, {\it Phys. Lett.} {\bf B294} (1992) 196.

\bibitem{dmw}
M.~J.~Duff, R.~Minasian and E.~Witten,
Nucl.\ Phys.\ B {\bf 465} (1996) 413
[hep-th/9601036];
G.~Aldazabal, A.~Font, L.~E.~Ibanez and F.~Quevedo,
Phys.\ Lett.\ B {\bf 380} (1996) 33
[hep-th/9602097];
N. Seiberg and E. Witten, {\it Nucl. Phys.} {\bf B471} (1996) 121
[hep-th/9603003].

\bibitem{FR}
P.G.O. Freund and M.A. Rubin, {\it Phys. Lett.} {\bf B97} (1980)
233.

\bibitem{Halliwell}
J.J. Halliwell, {\it Nucl. Phys.} {\bf B286} (1987) 729.

\bibitem{Witten}
E. Witten, Phys. Lett. {\bf B155} (1985) 151.

\bibitem{BFQ}
C.P. Burgess, A. Font and F. Quevedo, Nucl. Phys. {\bf B272}
(1986) 661.

\bibitem{Cremmer}
E. Cremmer, B. Julia, J. Scherk, S. Ferrara, L. Girardello and P.
van Nieuwenhuizen, {\it Nucl. Phys.} {\bf B147} (1979) 105.

\bibitem{BW}
E. Witten and J. Bagger, {\it Phys. Lett.} {\bf B115} (1982) 202.

\bibitem{RDSS}
S. Randjbar-Daemi, A. Salam and J. Strathdee, Nucl. Phys. {\bf
B214} (1983) 491.

\bibitem{SNRT}
M. Dine and N. Seiberg, {\it Phys. Rev. Lett.} {\bf 57} (1986)
2625.

\bibitem{FITerms}
M. Dine, N. Seiberg and E. Witten, {\it Nucl. Phys.} {\bf B289}
(1987) 589.

\bibitem{NRT}
M. Grisaru, M. Ro\v cek and W. Siegel, {\it Nucl. Phys.} {\bf B159} (1979) 429.

\bibitem{Wilsonvs1PI}
K.~A.~Intriligator and N.~Seiberg,
Nucl.\ Phys.\ Proc.\ Suppl.\  {\bf 45BC} (1996) 1
[hep-th/9509066];
M.~A.~Shifman,
Prog.\ Part.\ Nucl.\ Phys.\  {\bf 39} (1997) 1
[hep-th/9704114].

\bibitem{GCondensation}
C.P. Burgess, J.P. Derendinger and F. Quevedo and M. Quiros,, {\it
Ann. Phys.} {\bf 250} (1996) 193 hep-th/9505171;
 {\it Phys. Lett.} {\bf B348} (1995) 428, hep-th/9501065.

\bibitem{WbgVolII}
S. Weinberg, {\it The Quantum Theory of Fields II} Cambridge
University Press (1996).

\bibitem{WbgRG}
S.~Weinberg, ``Why The Renormalization Group Is A Good Thing,''
{\it  In *Cambridge 1981, Proceedings, Asymptotic Realms Of
Physics*, 1-19}.

\bibitem{BM}
For an application of these arguments in another context see: C.P.
Burgess and A. Marini, {\it Phys. Rev.} {\bf D45} (1992) 17.

\bibitem{Kstabilization}
J.A. Casas, {\it Nucl. Phys. Proc. Suppl.} {\bf 52A} (1997) 289
hep-th/9608010;\\
T. Banks and M. Dine, {\it Phys. Rev.} {\bf D50} (1994) 7454.

\bibitem{DS}
M.~Dine and N.~Seiberg,
Phys.\ Lett.\ B {\bf 162} (1985) 299.

\bibitem{racetrack}
N.V.~Krasnikov, Phys.~Lett. {\bf B193} (1987) 37.

\bibitem{WittenHeirarchy}
E. Witten, {\it Nucl. Phys.} {\bf B188} (1981) 513.

\bibitem{SoftTerms}
For a review with references see: A. Brignole, L.E. Ib\'a\~nez and
C. Mu\~noz, in {\it Perspectives on Supersymmetry}, ed. by G.L.
Kane, 1997, pp. 125  hep-ph/9707209.

\bibitem{sixd}
J.~W.~Chen, M.~A.~Luty and E.~Ponton,
JHEP {\bf 0009} (2000) 012
hep-th/0003067;
F.~Leblond, R.~C.~Myers and D.~J.~Winters,
JHEP {\bf 0107} (2001) 031
hep-th/0106140.
\bibitem{chris}
See for instance:
M.~Cvetic, H.~Lu and C.~N.~Pope,
Phys.\ Rev.\ Lett.\  {\bf 83} (1999) 5226
[hep-th/9906221];
M.~Cvetic, H.~Lu and C.~N.~Pope,
Nucl.\ Phys.\ B {\bf 597} (2001) 172
[hep-th/0007109]; M.~Cvetic, H.~Lu, C.~N.~Pope, A.~Sadrzadeh and T.~A.~Tran,
Nucl.\ Phys.\ B {\bf 590} (2000) 233
[hep-th/0005137], and references therein.
\bibitem{jan}
See for instance:
I.~Antoniadis, E.~Gava, K.~S.~Narain and T.~R.~Taylor,
Nucl.\ Phys.\ B {\bf 511} (1998) 611
[hep-th/9708075];
T.~R.~Taylor and C.~Vafa,
Phys.\ Lett.\ B {\bf 474} (2000) 130
[hep-th/9912152];
S.~Gukov, C.~Vafa and E.~Witten,
Nucl.\ Phys.\ B {\bf 584} (2000) 69
[Erratum-ibid.\ B {\bf 608} (2001) 477]
[hep-th/9906070];
P.~Mayr,
Nucl.\ Phys.\ B {\bf 593} (2001) 99
[hep-th/0003198];
G.~Curio, A.~Klemm, D.~Lust and S.~Theisen,
Nucl.\ Phys.\ B {\bf 609} (2001) 3
[arXiv:hep-th/0012213];
J.~Louis and A.~Micu,
Nucl.\ Phys.\ B {\bf 635} (2002) 395
[hep-th/0202168];
~D'Auria, S.~Ferrara and S.~Vaula,
New J.\ Phys.\  {\bf 4} (2002) 71
[hep-th/0206241];
L.~Andrianopoli, R.~D'Auria, S.~Ferrara and M.~A.~Lledo,
Nucl.\ Phys.\ B {\bf 640} (2002) 63
[hep-th/0204145].

\bibitem{ergin}
H.~Nishino and E.~Sezgin,
Phys.\ Lett.\ B {\bf 144} (1984) 187;
Nucl.\ Phys.\ B {\bf 278} (1986) 353;
Nucl.\ Phys.\ B {\bf 505} (1997) 497
[hep-th/9703075].
\bibitem{romans}
L.~J.~Romans,
Nucl.\ Phys.\ B {\bf 269} (1986) 691.
\bibitem{sethi}
K.~Dasgupta, G.~Rajesh and S.~Sethi,
JHEP {\bf 9908}, 023 (1999)
[hep-th/9908088].
\bibitem{jan2}
S.~Gurrieri, J.~Louis, A.~Micu and D.~Waldram,
[hep-th/0211102];
S.~Kachru, M.~B.~Schulz, P.~K.~Tripathy and S.~P.~Trivedi,
[hep-th/0211182].
\bibitem{curio}
G.~Curio, A.~Klemm, B.~Kors and D.~Lust,
Nucl.\ Phys.\ B {\bf 620}, 237 (2002)
[hep-th/0106155].
\end{thebibliography}
\end{document}